\documentclass{IEEEtran}
\usepackage{cite}
\usepackage{amsmath,amssymb,amsfonts}
\usepackage{algorithmic}
\usepackage{graphicx}
\usepackage{textcomp}
\usepackage{todonotes}

\def\BibTeX{{\rm B\kern-.05em{\sc i\kern-.025em b}\kern-.08em
    T\kern-.1667em\lower.7ex\hbox{E}\kern-.125emX}}
\begin{document}

\title{Time Series Data Cleaning with Regular and Irregular Time Intervals (Technical Report)}

\author{{Xi Wang}, {Chen Wang}\\
School of Software, Tsinghua University, Beijing 100084, China}

\maketitle

\begin{abstract}
Errors are prevalent in time series data, especially in the industrial field. 
Data with errors could not be stored in the database, which results in the loss of data assets.
Handling the dirty data in time series is non-trivial, when given irregular time intervals.
At present, to deal with these time series containing errors, besides keeping original erroneous data, discarding erroneous data and manually checking erroneous data, we can also use the cleaning algorithm widely used in the database to automatically clean the time series data.
This survey provides a classification of time series data cleaning techniques and comprehensively reviews the state-of-the-art methods of each type. 
In particular, we have a special focus on the irregular time intervals.
Besides we summarize data cleaning tools, systems and evaluation criteria from research and industry. 
Finally, we highlight possible directions time series data cleaning.
\end{abstract}

\section{Introduction}
\label{sec:introduction}

\par
Time series data can be defined \cite{shumway2017time} as a sequence of random variables, $x_1$, $x_2$,..., $x_n$, where the random variable $x_1$ denotes the value taken by the series at the first time point, the variable $x_2$ denotes the value for the second time period, $x_n$ denotes the value for the n-th time period, and so on. Time series have been widely used in many fields \cite{hamilton1994time,brockwell2016introduction,box2015time} such as financial economy, meteorology and hydrology, signal processing, industrial manufacturing, and so on. Time series data are important in industry, where there are all kinds of sensor devices capturing data from the industrial environment  uninterruptedly.
Owing to the fact that data of the sensor devices are often unreliable \cite{DBLP:conf/pervasive/JefferyAFHW06}, time series data are often large and dirty.
 In the financial field, the most important application of time series data is to predict future commodity (stock) price movements.
 However, time series errors in the financial field are also very prevalent, even some data sets, which are considered quite accurate, still have erroneous data.
For instance, the correct rate of stock information on Yahoo Finance is 93\%.
The costs and risks of errors, conflicts, and inconsistencies in time series have drawn widespread attention from businesses and government agencies.
In recent work, the data quality issues in time series data are studied, since they pose unique data quality challenges due to the presence of autocorrelations, trends, seasonality, and gaps in the time series data \cite{DBLP:journals/debu/DasuDS16}.
According to Shilakes et al. \cite{shilakes1998enterprise}, the relevant market growth rate of data quality is about 17\%, which is much higher than the 7\% annual growth rate of the IT industry. For instance, approximately 30\% to 80\% of the time and cost are spent on data cleaning in data warehousing project development.
The time series errors can be either timestamp errors \cite{DBLP:journals/pvldb/SongC016} or observed value errors. 
In this survey, we focus on the existing methods of dealing with observed value errors, thereby, the time series errors mentioned in the following are observation errors. There are two types of processing methods commonly used in the industry when dealing with time series data errors:
\par(1) Discarding erroneous data. First, the time series is detected via using an anomaly detection algorithm, and then the detected abnormal data are discarded.
\par(2) Cleaning data. Data cleaning is divided into manual cleaning and automatic cleaning. There is no doubt that manual cleaning has a high accuracy rate, but it is difficult to implement because it takes more time and effort.

The existing surveys of data cleaning \cite{DBLP:conf/sigmod/ChuIKW16,hellerstein2008quantitative} mainly summarize the methods of dealing with data missing \cite{DBLP:journals/pvldb/KhayatiLTC20}, data inconsistence, data integration and erroneous data in the database. Karkouch et al. \cite{DBLP:journals/jnca/KarkouchMMN16} review the generation of sensor data, the reasons for the formation of data quality problems, and the techniques for improving data quality. However, Karkouch et al. \cite{DBLP:journals/jnca/KarkouchMMN16} do not provide a detailed overview of the existing state-of-the-art of erroneous data cleaning. Thereby,\begin{bfseries} we review the state-of-the-art of time series data error value cleaning\end{bfseries}, which may provide a tutorial for others.

\subsection{Problem Statement}
\begin{figure}
\includegraphics[width=3.3 in]{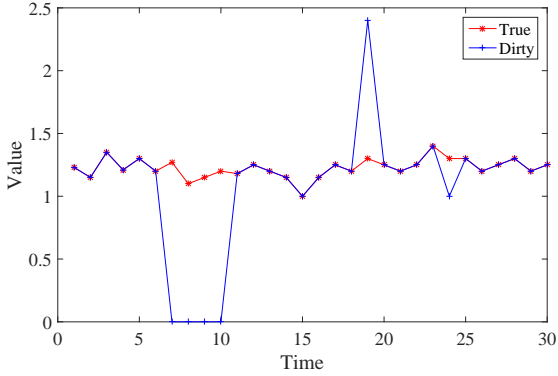}
\caption{An Example Error Type}
\label{error_type}
\end{figure}

In this study, enlightened by related research \cite{tsay1988outliers,DBLP:journals/tkde/GuptaGAH14} and \cite{zhangaoqian} on the classification of time series error types, we summarize the common error in time series into three categories, namely, single point big error, single point small error and continuous errors.
This article takes the stock price of a stock for 30 consecutive trading days as an example. As shown in Figure \ref{error_type}, the characteristics of these three types of errors are described in detail. The red line in the figure indicates the true price of the stock in 30 consecutive trading days, and the blue line indicates the price of the stock crawled by a website. For various reasons, the observed value may not be the same as the actual value. It can be seen that in the four consecutive trading days of 8-11, the observed values are all 0, and the true values are 1.3, 1.2, 1.1 and 1.15, respectively, on the 20th trading day, the observed value is 2.4 and the true value is 1.3, on the 25th trading day, the observed value is 1 and the true value is 1.3.

\par (1) Continuous errors. The so-called continuous errors, that is, in the time series, errors occur in several consecutive time points. Specifically, continuous errors can continue to be subdivided into several types \cite{tsay1988outliers}, but no longer detailed here. The observed values from the 8th to 11th trading days in Figure \ref{error_type} are all 0, that is, continuous errors occur here. Continuous errors are common in real life. For instance, when someone is holding a smartphone and walking on the road, nearby tall buildings may have a lasting impact on the collected GPS information. Besides, system errors can also cause continuous errors.

\par (2) A single big error. A single point error is an error that occurs discontinuously in a time series and only occurs on a single data point at intervals. A big error means that the observed value of the data point is far from the true value. Remarkably, the size of the error is relative and closely related to the real situation of the data set. As shown on the 20th trading day in Figure \ref{error_type}, the observed value differs from the true value by 1.1. Compared with the 25th trading day when the observed value differs from the true value by 0.3, the error of this data point is large, so this data point error is a single big error. The single point big error is also very common in daily life. For instance, the data of motor vehicle oil level recorded by cursors may cause a single big error when bumping on the road.
 
\par (3) A single small error. Similar to a single point big error, that is, errors do not occur consecutively, only on a single data point at intervals. When the observed value of the data point differs from the true value by a small distance, on the 25th trading day in Figure \ref{error_type}, it is a single point small error. As stated in \cite{DBLP:conf/sigmod/BohannonFFR05}, the rationale behind single-point small errors is that people or systems always try to minimize possible errors. For instance, people may only have some small omissions when copying files.

\par
\begin{figure}
\includegraphics[width=3.3 in]{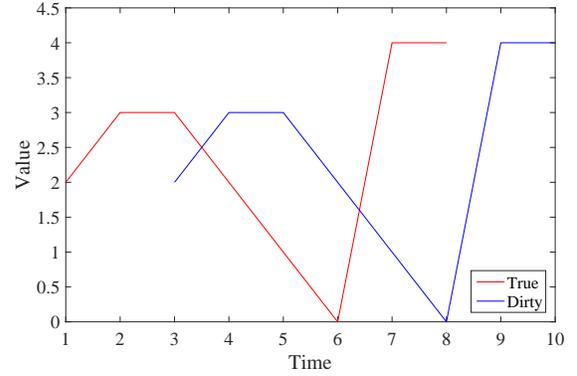}
\caption{An Example of Translational Error}
\label{Translational-error}
\end{figure}

 (4) Translational error. As shown in Figure \ref{Translational-error}, where $x$ axis represents time and $y$ axis represents the value of the corresponding time, the red line represents true value, and the blue line represents error value after the translation, the solution to this type of error is not as much as mentioned above.

\par Ignoring time series errors often results in unpredictable consequences for a series of applications such as query analysis. Thereby, time series cleaning algorithms are very important for mining the potential value of data. This paper reviews the cleaning algorithm and anomaly detection algorithm of time series data. By summarizing the existing methods, a reference or guidance is given to scholars interested in time series data cleaning and based on this, the possible challenges and future work of time series cleaning topics are discussed.

\subsection{Problem Challenge}
For the problem of time series data cleaning, the following four difficulties have been discovered through the survey:

\par (1) The amount of data is large and the error rate is high. The main source of time series data is sensor acquisition. Especially in the industrial field, sensors distributed throughout the machine are constantly monitoring the operation of the machine in real-time.
These sensors often collect data at a frequency of seconds, and the amount of data collected is quite large. For instance, the sensor collection interval of a wind power company equipment is 7 seconds, each machine has more than 2000 working status data, and more than 30 million pieces of data are collected every day, so the working status data of one day could exceed 60 billion.
However, the data collected by the sensor are often not accurate enough, and some because of the physical quantity of the observation is difficult to measure accurately. For instance, in a steel mill, with affecting by environmental disturbances the surface temperature of the continuous casting slab cannot be accurately measured or may cause distortion due to the power of the sensor itself.

\par (2) The reasons for generating time series data errors are complicated.
People always try to avoid the generation of time series erroneous data, however, there are various time series errors. Besides the observed errors that we mentioned above for various reasons, Karkouch et al. \cite{DBLP:journals/jnca/KarkouchMMN16} also explain in detail the IoT data errors generated by various complex environments. 
IoT data is a common time series of data, and its widespread existence is really in the world. The complex reasons of time series errors also are challenges we face in cleaning and analyzing data that is different from traditional relational data. 

\par (3) Time series data are continuously generated and stored.
The biggest difference between time series data and relational data is that the time series is continuous.
Thereby, for time series data, it is important that the cleaning algorithm supports online operations (real-time operations).
The online anomaly detection or cleaning algorithm can monitor the physical quantity in real-time, detect the problem and then promptly alarm or perform a reasonable cleaning. Thereby, the time series cleaning algorithm is not only required to support online calculation or streaming calculation but also has good throughput. 

\par (4) Minimum modification principle \cite{DBLP:conf/icdt/AfratiK09,DBLP:journals/iandc/ChomickiM05,DBLP:conf/pods/FaginKK15}.
Time series data often contain many errors. Most of the widely used time series cleaning methods utilize the principle of smooth filtering. Such methods may change the original data too much, and result in the loss of the information contained in the original data.
Data cleaning needs to avoid changing the original correct data. It should be based on the principle of minimum modification, that is, the smaller the change, the better.

\subsection{Organization}
Different algorithms tackle these challenges in different ways, which usually include smoothing-based methods, constraint-based methods, and statistical-based methods as shown in Table \ref{overview}. Besides some time series anomaly detection algorithms can also be effectively used to clean data. The remainder of this paper is organized as follows. The aforesaid four types of algorithms are discussed from Section \ref{sec:Smoothing based cleaning algorithm} to \ref{sec:Time series anomaly detection}, respectively. In Section \ref{sec:Tools and Evaluation criteria} we introduce existing time series cleaning tools, systems, and evaluation criteria. Finally, we summarize this paper in Section \ref{sec:Conclusion and Future Directions} and discuss possible future directions.

\begin{table}
 \centering\small
\caption{The Overview of Methods}
\label{overview}
\setlength{\tabcolsep}{3pt}
\begin{tabular}{|p{70pt}|p{150pt}|}
\hline
 Type& Method\\\hline
Smoothing based (Section \ref{sec:Smoothing based cleaning algorithm})& Moving Average \cite{DBLP:books/daglib/0005327}\par AutoRegressive \cite{box2015time,DBLP:journals/envsoft/HillM10,DBLP:conf/kdd/YamanishiT02}\par Autoregressive Moving
 Average Model \cite{dilling2017cleaning,tsay1988outliers,DBLP:conf/icassp/AlengrinF78}\par Kalman Filter \cite{kalman1960new,marczak2018data,morrison1977kalman,brown1992introduction,DBLP:journals/tsp/EinickeW99,DBLP:journals/taslp/GohTT99,DBLP:conf/icdcs/Zhuang0WL07,gardner2006exponential}\par Interpolation \cite{DBLP:conf/icdm/KeoghCHP01,xu2015data} \par State-space Model \cite{jones1966exponential,van2005accurate,marczak2018data}
 \par Trajectory Simplification\cite{DBLP:journals/tip/ChenXF12,DBLP:journals/pvldb/LongWJ14} \\
\hline
Constraint based (Section \ref{sec:Constraint based cleaning algorithm})&  Order Dependencies \cite{DBLP:conf/sigmod/DongH82,DBLP:journals/tcs/GinsburgH83,DBLP:conf/pods/GinsburgH83,DBLP:journals/jacm/GinsburgH86}\par Denial Constraints \cite{DBLP:conf/icde/LopatenkoB07}\par Sequential Dependencies \cite{DBLP:journals/pvldb/GolabKKSS09}
\par Speed Constraints \cite{DBLP:conf/sigmod/SongZWY15}\par Variance Constraints \cite{DBLP:conf/dasfaa/YinYWHL18}
\par Similarity Rule Constraints  \cite{DBLP:journals/tkde/SongSZCW20}
\par Learning Individual Models \cite{DBLP:conf/icde/ZhangSSW19} 
\par Temporal Dependence \cite{DBLP:journals/pvldb/CaiX0JOZ18,DBLP:conf/kdd/ChengB014}\\
\hline
Statistics based (Section \ref{sec:Statistics based cleaning algorithm})&  Maximum Likelihood \cite{DBLP:journals/tsp/BreslerM86,DBLP:conf/icdt/GogaczT17,DBLP:journals/tosn/WangKA14,DBLP:conf/sigmod/YakoutBE13,DBLP:conf/sigmod/ZhangSW16}\par Bayesian Model \cite{DBLP:conf/sigmod/WangDM15,DBLP:conf/icml/GetoorFKT01}\par Markov Model \cite{dukhovny1900markov,cai1994markov,DBLP:journals/tcom/ZhangK99}\par Hidden Markov Model \cite{DBLP:journals/eswa/HassanNK07,DBLP:journals/eswa/DongYKHEK09,gupta2012stock,DBLP:conf/sigmod/BabaJLPKX16}
\par SMURF \cite{leema2011effective,DBLP:journals/ijguc/XuDLSW18,DBLP:conf/vldb/JefferyGF06}
\par Spatio-Temporal Probabilistic Model \cite{DBLP:conf/icde/ZhengMC19}\par Expectation-Maximization \cite{shumway1982approach} \par Relationship-dependent Network \cite{DBLP:conf/sigmod/BergmanMNT15,DBLP:journals/jmlr/NevilleJ07,DBLP:conf/sigmod/MayfieldNP10}\\
\hline
Anomaly detection (Section \ref{sec:Time series anomaly detection}) & Density-Based Spatial Clustering of Applications with Noise \cite{DBLP:conf/cyberc/DiaoLMYH15,DBLP:journals/bdr/CorizzoCJ19} \par Local Outlier Factor \cite{DBLP:conf/cyberc/DiaoLMYH15}\par Abnormal Sequence Detection \cite{DBLP:journals/kais/KeoghLLH07,DBLP:journals/gpem/GonzalezD03,dasgupta2002anomaly,DBLP:conf/icde/AggarwalY15}\par Window-based Anomaly Detection \cite {DBLP:conf/icde/KontakiGPTM11}\par Generative Adversarial Networks \cite{DBLP:conf/icann/LiCJSGN19,DBLP:conf/itsc/SunPLS18,DBLP:conf/cikm/FangSCG19}\par Long Short-Term Memory \cite{DBLP:journals/corr/FilonovLV16,DBLP:journals/corr/MalhotraTRAVAS16,DBLP:conf/esann/MalhotraVSA15} \par Speed-based cleaning algorithm \cite{wangxi}\\

\hline
\end{tabular}
\end{table}

\section{Smoothing based cleaning algorithm}
\label{sec:Smoothing based cleaning algorithm}
Smoothing techniques are often used to eliminate data noise, especially numerical data noise. Low-pass filtering, which filters out the lower frequency of the data set, is a simple algorithm. The characteristic of this type of technology is that the time overhead is small, but because the original data may be modified much, which makes the data distorted and leads to the uncertainty of the analysis results, there are not many applications used in time series cleaning. The research of smoothing technology mainly focuses on algorithms such as Moving Average (MA) \cite{DBLP:books/daglib/0005327}, Autoregressive (AR) \cite{box2015time,DBLP:journals/envsoft/HillM10,DBLP:conf/kdd/YamanishiT02} and Kalman filter model \cite{kalman1960new,marczak2018data,morrison1977kalman}. Thereby, this chapter mainly introduce these three technologies and their extensions.

\subsection{Moving average}
The moving average (MA) series algorithm \cite{DBLP:books/daglib/0005327} is widely used in time series for smoothing and time series prediction. A simple moving average (SMA) algorithm: Calculate the average of the most recently observed $N$ time series values, which is used to predict the value at time $t$. A simple definition as shown in equation \eqref{ma}.
\begin{equation}\widehat{x_t}=\cfrac{1}{2n+1}\sum^n_{i=-n}x_{i+t}\label{ma}\end{equation}
\par In equation \eqref{ma}, $\widehat{x_t}$ is the predicted value of $x_t$, $x_t$ represents the true value at time $t$, $2*n+1$ is the window size(count). For a time series $x(t)$={9.8, 8.5, 5.4, 5.6, 5.9, 9.2, 7.4}, for simplicity, we use SMA with $n = 3$ and $k = 4$, then calculate according to equation \eqref{sma}.
\begin{equation}x_t=\cfrac{(x_{t-n}+x_{t-2}...+x_{t+n})}{(2n+1)}=7.4\label{sma}\end{equation}
\par To eliminate errors or noises, the data $x(t)$ can be considered as a certain time window of sliding window, then continuously calculate the local average over a given interval $2n+1$ to filter out the noise (erroneous data) and get a smoother measurement.
\par In the weighted moving average (WMA) algorithm, data points at different relative positions in the window have different weights. Generally defined as:
\begin{equation}\widehat{x_t}=\sum^n_{i=-n}\omega_ix_{i+t}\label{wma}\end{equation}
\par
In equation \eqref{wma}, $\omega_i$ represents the weight of the influence of the $i$ position data point on the $t$ position data point, other definitions follow the example above.
A simple strategy is that the farther away from the two data points, the smaller the mutual influence. For instance, a natural idea is the reciprocal of the distance between two data points as the weight of their mutual influence. Similarly, the weight of each data point in the exponential weighted moving average (EWMA) algorithm \cite{gardner2006exponential} decreases exponentially with increasing distance, which is mainly used for unsteady time series \cite{brown2004smoothing,holt2004forecasting}.
\par Aiming at the need for the rapid response of sensor data cleaning, Zhuang et al. \cite{DBLP:conf/icdcs/Zhuang0WL07} propose an intelligent weighted moving average algorithm, which calculates weighted moving averages via collected confidence data from sensors. Zhang et al. \cite{DBLP:journals/tits/ZhangYZHL13} propose a method based on multi-threshold control and  approximate positive transform to clean the probe vehicle data, and fill the lost data with the weighted average method and exponential smoothing method. Qu et al. \cite {qu2016data} first use cluster-based methods for anomaly detection and then use exponentially weighted averaging for data repair, which is used to clean power data in a distributed environment.

\subsection{Autoregressive}
The Autoregressive (AR) Model is a process that uses itself as a regression variable and uses the linear combination of the previous $k$ random variables to describe the linear regression model of the random variable at the time $t$. The definition of AR model \cite{DBLP:conf/icde/VolkovsCSM14,DBLP:journals/envsoft/HillM10} as shown in equation \eqref{ar}.
\begin{equation}\widehat{x_t}=\sum^k_{i=1}\omega_{i}x_{t-i}+\epsilon_t+a\label{ar}\end{equation}
\par In equation \eqref{ar}, $\widehat{x_t}$ is the predicted value of $x_t$, $x_t$ represents the true value at time $t$, $k$ is the order, $\mu$ is mean value of the process, $\epsilon_t$ is white noise, $\omega_{i}$ is the parameter of the model, $a$ is a constant.
\par Park et al. \cite{park2005outlier} use labeled data $y$ to propose an autoregressive with exogenous input (ARX) model based on the AR model:
\begin{equation}\widehat{y_t}=x_t+\sum^k_{i=1}\omega_{i}(y_{t-i}-x_{t-i})+\epsilon_t\label{arx}\end{equation}
\par In equation \eqref{arx}, $\widehat{y_t}$ is the possible repair of $x_t$, and others are the same to the aforesaid AR model.
Alengrin et al. \cite{DBLP:conf/icassp/AlengrinF78} propose Autoregressive moving average (ARMA) model), which is composed of the AR model and MA model. Besides that, the Gaussian Autoregressive Moving Average model is defined as shown in equation \eqref{arma} \cite{tsay1988outliers}.
\begin{equation}\Phi(B)Z_t=\theta_0+\theta(B)x_t\label{arma}\end{equation}
\par 
In equation \eqref{arma}, $\Phi(B)=1-\Phi_1B-...-\Phi_pB^p$ and $\theta(B)=1-\theta_1B-...-\theta_qB^q$ are polynomial in $B$ of degrees $p$ and $q$, respectively, $\theta_0$ is a constant, $B$ is the backshift operator such that $BZ_t=Z_{t-1}$, and {$x_t$} is a sequence of independent Gaussian variates with mean $\mu$=0 and variance $\sigma_x^2$. 
Box et al. \cite{box1970distribution} propose a more complex Autoregressive Integrated Moving Average (ARIMA) model based on the ARMA model, which is not described in detail here. Akouemo et al. \cite{akouemo2017data} propose a method combining ARX and Artificial Neural Network (ANN) model for cleaning time series, which performs a hypothesis test to detect anomalies the extrema of the residuals, and repairs anomalous data points by using the ARX and ANN models. Dilling et al. \cite{dilling2017cleaning} clean high-frequency velocity profile data with ARMA model and Chen et al. \cite{chen2009arima} use the ARIMA model to clean wind power data.

\subsection{Kalman filter model}

Kalman \cite{kalman1960new} proposes the Kalman filter theory, which can deal with time-varying systems, non-stationary signals, and multi-dimensional signals. Kalman filter creatively incorporates errors (predictive and measurement errors) into the calculation, the errors exist independently and are not affected by measured data. The Kalman model involves probability, random variable, Gaussian Distribution, and State-space Model, etc. Consider that the Kalman model involves too much content, no specific description is given here, and only a simple definition is given. First, we introduce a system of discrete control processes which can be described by a Linear Stochastic Difference equation as shown in equation \eqref{kalman-0}.
\begin{equation}{x_t}=mx_{t-1}+nv_t+p_t\label{kalman-0}\end{equation}
Also, the measured values of the system are expressed as shown in equation \eqref{kalman-2}.
\begin{equation}{y_t}=rx_{t-1}+q_t\label{kalman-2}\end{equation}
\par In equation \eqref{kalman-0} and \eqref{kalman-2}, $x_t$ is the system state value at time $t$, and $v_t$ is the control variable value for the system at time $t$. $m$ and $n$ are system parameters, and for multi-model systems, they are matrices, $y_t$ is the measured value at time $t$, $r$ is the parameter of the measurement system, and for multi-measurement systems, $r$ is a matrix, $p(k)$ and $q(k)$ represent the noises of the process and measurement, respectively, and they are assumed to be white Gaussian Noise.

\par The extended Kalman filter is the most widely used estimation for a recursive nonlinear system because it simply linearizes the nonlinear system models. However, the extended Kalman filter has two drawbacks: linearization can produce unstable filters and it is hard to implement the derivation of the Jacobian matrices. Thereby, Ma et al. \cite{ma2004predict} present a new method of predicting the Mackey-Glass equation based on the unscented Kalman filter to solve these problems. In the field of signal processing, there are many works \cite{brown1992introduction,DBLP:journals/tsp/EinickeW99,DBLP:journals/taslp/GohTT99} based on Kalman filtering, but these techniques have not been widely used in the field of time series cleaning. Gardner et al. \cite{gardner2006exponential} propose a new model, which is based on the Kernel Kalman Filter, to perform various nonlinear time series processing. Zhuang et al. \cite{DBLP:conf/icdcs/Zhuang0WL07} use the Kalman filter model to predict sensor data and smoothed it with WMA.
\subsection{Trajectory Simplification}
The main purpose of trajectory simplification is to reduce the original trajectory from $N$ trajectory data points to $M$ trajectory data points, $M<N$. And retain important location or topological features. The compression algorithm needs to ensure that the sequence of $M$ points has a minimum compression error ($M$-$\epsilon$ problem), Such as MRPA\cite{DBLP:journals/tip/ChenXF12}, Error-Search\cite{DBLP:journals/pvldb/LongWJ14}. The input of the Error-Search algorithm is the target compression ratio $\lambda$, Original track $T$. Returns the compression that satisfies the given compression ratio $\lambda$ with the smallest DAD error. The main steps of the Error-Search algorithm are as follows: (1) According to the original trajectory $T$, Build search space $\epsilon$, create a search $\epsilon$ space based on the opposite direction in a more efficient way
Trace $T$. (2) Error checking. the main task is to determine whether there is a simplified trajectory $\hat{T}$, satisfy $DAD(\hat{T})<\epsilon$. Error-Search is designed to retain the direction information of the track. The time complexity of the algorithm is $O(N^{2}\log N)$, the space complexity is  $O(N^{2})$. Error-Search is an accurate trajectory simplification algorithm that retains direction information, but the time complexity is too high. \cite{DBLP:journals/pvldb/LongWJ14} proposed an approximate algorithm Span-Search algorithm. Span-Search algorithm uses approximate error to replace time complexity $O(N\log^{2}N)$ space complexity $O(N)$.
\subsection{Summary and Discussion}
As shown in Table \ref{Summary of smoothing}, there are many methods based on smoothing, such as the state-space model \cite{jones1966exponential,van2005accurate,marczak2018data} and Interpolation \cite{DBLP:conf/icdm/KeoghCHP01,xu2015data}. The state-space model assumes that the system's change over time can be determined by an unobservable vector sequence, the relationship between the time series and the observable sequence can be determined by the state-space model. By establishing state equations and observation equations, the state-space model provides a model framework to fully describe the temporal characteristics of dynamic systems. To make this kind of smoothing algorithm have a better effect, many studies \cite{chang1988estimation,DBLP:journals/tsp/SwamiM90,plett2004extended} have also proposed various techniques to estimate the parameters in the above methods. Most smoothing techniques, when cleaning time series, have a small-time overhead, but it is very easy to change the original correct data points, which greatly affects the accuracy of cleaning. In other words, correct data are altered, which can distort the results of the analysis and lead to uncertainty in the results. 

\begin{table}
 \centering\small
\caption{Summary of Smoothing}
\label{Summary of smoothing}
\setlength{\tabcolsep}{3pt}
\begin{tabular}{|p{70pt}|p{150pt}|}
\hline
 Reference& Method\\ \hline
\cite{DBLP:books/daglib/0005327} & MA\\
\cite{DBLP:journals/tits/ZhangYZHL13,DBLP:conf/icdcs/Zhuang0WL07} & WMA\\
\cite{gardner2006exponential,DBLP:journals/tits/ZhangYZHL13,qu2016data} & EWMA\\
\cite{DBLP:journals/envsoft/HillM10,DBLP:conf/kdd/YamanishiT02,DBLP:conf/icde/VolkovsCSM14} &AR\\ 
\cite{akouemo2017data,park2005outlier} & ARX\\
\cite{dilling2017cleaning,tsay1988outliers,DBLP:conf/icassp/AlengrinF78} & ARMA\\
\cite{chen2009arima,box1970distribution} & ARIMA\\
\cite{kalman1960new,marczak2018data,morrison1977kalman}\par \cite{brown1992introduction,DBLP:journals/tsp/EinickeW99,DBLP:journals/taslp/GohTT99}\par
\cite{DBLP:conf/icdcs/Zhuang0WL07,gardner2006exponential}&Kalman Filter Model\\
\cite{ma2004predict}&The Unscented Kalman Filter\\
\cite{jones1966exponential,van2005accurate,marczak2018data} & The State-space Model\\
\cite{DBLP:conf/icdm/KeoghCHP01,xu2015data} & Interpolation \\
\cite{chang1988estimation,DBLP:journals/tsp/SwamiM90,plett2004extended} & To Estimate the Parameters of Model\\
\hline
\end{tabular}
\end{table}

\section{Constraint based cleaning algorithm}
\label{sec:Constraint based cleaning algorithm}
In this section, we introduce several typical algorithms, which include order dependencies (ODs) \cite{DBLP:conf/sigmod/DongH82}, sequential dependencies (SDs) \cite{DBLP:journals/pvldb/GolabKKSS09} and speed constraints \cite{DBLP:conf/sigmod/SongZWY15}, for repairing time series errors. 

\subsection{Order Dependencies}

In relational databases, Order Dependencies (ODs) are simple and effective methods, which have been widely studied \cite{DBLP:conf/sigmod/DongH82,DBLP:journals/tcs/GinsburgH83,DBLP:conf/pods/GinsburgH83,DBLP:journals/jacm/GinsburgH86}. We find that ODs are also suitable for solving some time series data cleaning problems. The specific explanation is as follows: Let $x(t)={x_1,x_2...x_t}$ be a time series, ODs can be expressed by $<,\leq,>,\geq$. For the number of miles traveled by the car $x(t)$, the mileage should increase over time. Formal representation is as follows:
$t_1<t_2$ then $x_{t_1}\leq x_{t_2}$ where $x(t)$ is mileage, $t$ is timestamp. For instance, consider an example relation instance in Table \ref{table-OD}.
The tuples are sorted on attribute $\textsf{sequence number}$, which identifies sea level that rapidly increase from hour to hour.

\begin{table}
 \centering\small
\caption{An Example of Order Dependencies}
\label{table-OD}
\setlength{\tabcolsep}{3pt}
\begin{tabular}{|p{25pt}|p{80pt}|p{35pt}|}
\hline
& sequence number& time\\\hline
$t_1$ & 1 & 1\\
 $t_2$ & 2 & 4 \\
 $t_3$ & 3 & 10 \\
 $t_4$ & 4 & 13 \\
 $t_5$ & 5 & 17 \\
 $t_6$ & 6 & 21 \\
 $t_7$ & 7 & 23 \\
\hline
\end{tabular}
\end{table}

\par Generally, ODs in the form of equation \eqref{ODs} states that 
$N$ is strictly increasing with $M$. Such as equation \eqref{ODs2}.
\begin{equation}M\rightarrow_{(0,\infty)} N\label{ODs}\end{equation}
\begin{equation}\mathsf{hour}\rightarrow_{(0,\infty)}\mathsf{height}\label{ODs2}\end{equation}
\par ODs and DCs can also be used as an integrity constraint for error detection and data repairing in databases. Wijsen \cite{DBLP:journals/is/Wijsen98,DBLP:journals/tkde/Wijsen01} extends ODs with a time dimension for temporal databases. Let $I=\{I_1, I_2, I_3, \dots \}$ be a temporal relation, which can be viewed as a time series  of conventional ``snapshot'' relations, all over the same set of attributes.
A trend dependency (TD) allows attributes with linearly ordered domains to be compared over time by using any operator of $\{ <, =, >, \leq, \geq, \neq \}$. Consider the constraint is specified over $(I_i, I_{i+1})$ in $I$. For each time point $i$, it requires comparing employee records at time $i$ with records at the next time $i+1$, such that salaries of employees should never decrease.
Lopatenko et al. \cite{DBLP:conf/icde/LopatenkoB07} propose a numerical type data cleaning method based on Denial Constraints (DCs) as constraints, whose principle is similar to this one.
It is also notable that the given denial constraints could also be dirty and should be cleaned together with data \cite{DBLP:conf/sigmod/SongZW16}.

\subsection{Sequential Dependencies}
\label{sect-SD}

The sequential dependency algorithm proposed by Golab et al. \cite{DBLP:journals/pvldb/GolabKKSS09} focuses on the difference in values between two consecutive data points in a time series. Golab et al. \cite{DBLP:journals/pvldb/GolabKKSS09} define the CSD Tableau Discovery Problem as
given a relation instance and an embedded SD $M\rightarrow_g N$, to find a tableau $t_r$ of minimum size such that the CSD $(M\rightarrow_g N, t_r)$ has confidence at least a given threshold. A CSD\ can be $$(\mathsf{hour}\rightarrow_{(0,\infty)}\mathsf{height}, \textnormal{[1961.01.01 00:00--2016.01.01 00:00]}).$$
It states that for any two consecutive hours in [1961.01.01 00:00--2016.01.01 00:00], their distance should always be $>0$.

\par Generally, a sequential dependency (SD) is in the form of 
\begin{equation}M\rightarrow_g N.\label{SD}\end{equation}
\par In equation \eqref{SD}, $M\subseteq\mathit{R}$ are ordered attributes, $N\subseteq\mathit{R}$ can be measured by certain distance metrics, and $g$ is an interval.
It states that when tuples are sorted on $M$, the distance between the $N$-values of any two consecutive tuples are within interval $g$.
Fischer et al. \cite{DBLP:conf/sensys/Casado-VaraPPC18} propose the concept of streaming mode to represent the structure and semantic constraints of data streams. The concept contains a variety of semantic information, including not only numeric values, but also attributes between order. The sequential dependency algorithm can be used not only for traditional relational database cleaning, but also for time series cleaning. 
In fact, there are many dependency-based cleaning algorithms designed for relational databases that are not suitable for time series data cleaning, such as: Functional Dependencies \cite {beeri1984structure} (FDs), Conditional Functional Dependencies \cite{fan2008conditional,DBLP:journals/tkde/FanGLX11} (CFDs),
Matching Dependencies (MDs) \cite{DBLP:conf/cikm/SongC09,DBLP:journals/dke/Song013,DBLP:journals/tkdd/WangSCYC17},
Differential Dependencies (DDs) \cite{DBLP:journals/tods/Song011,DBLP:conf/icde/SongCC12,DBLP:journals/tkde/Song0C14} or
Comparable Dependencies (CDs) \cite{DBLP:conf/icde/SongCY11,DBLP:journals/vldb/Song0Y13}. 
The sequential dependency is one of the few algorithms based on dependency that can be used for time series cleaning.

\subsection{Speed Constraints}
To clean time series data, speed constraint-based method \cite{DBLP:conf/sigmod/SongZWY15} considers the restrictions of speed on value changes in a given interval. As we have learned some common sense, e.g., the maximum flying speed of a bird, temperatures in a day, car mileage, etc. Consider with time window  $\mathit{T}$ is a pair of minimum speed $\mathit{S}_{\min}$ and maximum speed $\mathit{S}_{\max}$
over the time series  $\mathit{x} = \mathit{x}_1, \mathit{x}_2, \dots, \mathit{x}_t$, where each $\mathit{x}_i$ is the value of the $i$-th data point, with a timestamp $\mathit i$. \par For instance, consider time series:
\par $\mathit{x(t)} = \{150, 160, 170, 180, 110, 200, 210, 220, 230\}$
where timestamps $\mathit{t} = \{1,2,3,4,5,6,7,8,9\}$. 
The \textsf{value} attribute corresponds to $\mathit{x}$, while the \textsf{time} attribute in Table \ref{table-example-speed} denotes the timestamps.

\begin{table}
 \centering\small
\caption{An Example Relation Instance of Time Series}
\label{table-example-speed}
\setlength{\tabcolsep}{3pt}
\begin{tabular}{|p{25pt}|p{80pt}|p{35pt}|}
\hline
& time& value\\\hline
 $t_1$ & 1 & 150 \\
 $t_2$ & 2 & 160 \\
 $t_3$ & 3 & 170 \\
 $t_4$ & 4 & 180 \\
 $t_5$ & 5 & 110 \\
 $t_6$ & 6 & 200 \\
 $t_7$ & 7 & 210 \\
 $t_8$ & 8 & 220 \\
 $t_9$ & 9 & 230 \\
\hline
\end{tabular}
\end{table}

\begin{figure}
\includegraphics[width=3.3 in]{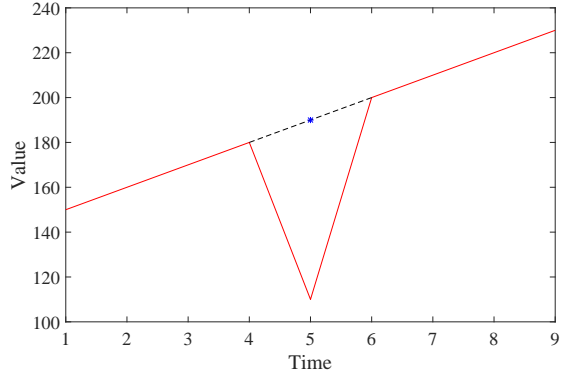}
\caption{An Example of Speed Constraints}
\label{fig-speed}
\end{figure}

Suppose a window size $\mathit{T} = 2$, $\mathit{S}_{\min} = -50$, and $\mathit{S}_{\max} = 50$ in the speed constraints, for data points $\mathit{x}_5$ and $\mathit{x}_4$, 
$\frac{110-180}{5-4}=-80<-50$.
Similarly, $\mathit{x}_5$ and $\mathit{x}_6$ with speed 
$\frac{200-110}{6-5}=90>50$ are violations to $\mathit{s}_{\max} = 50$. To remedy the violations (denoted by red lines), a repair on $\mathit{x}_5$ can be performed, i.e., $\mathit{x}'_5=190$, which is represented by the blue ``$*$'' symbol in Figure \ref{fig-speed}. As illustrated in Figure \ref{fig-speed}, the repaired sequence satisfies both the maximum and minimum speed constraints.

\par Generally, a speed constraint is in the form of equation \eqref{speed}.
\begin{equation}\mathit{S} = (\mathit{S}_{\min}, \mathit{S}_{\max})\label{speed}\end{equation}
\par If time series data $\mathit{x}$ satisfies the speed constraint $\mathit{S}$, then for any $\mathit{x}_i, \mathit{x}_j$ in a time window $\mathit{T}$, it has $\mathit{S}_{\min} < \frac{\mathit{x}_j - \mathit{x}_i}{\mathit j - \mathit i} < \mathit{S}_{\max}$.  In practical applications, speed constraints are often valid for a specific period of time. For instance, when considering the fastest speed of a car, the time period of the reference is often in hours, and two data points in different years are not considered. The value of the speed constraints $\mathit{S}$ may be positive (the growth rate of the constraint value) or negative (the rate of decline of the constraint value). Speed constraints are less effective when dealing with small errors, and Wei Yin et al. \cite{DBLP:conf/dasfaa/YinYWHL18} propose a further study of variance constraints, which use the variance threshold $V$ to measure the degree of dispersion of the time series in a given $W$ window.

\subsection{Temporal Dependence}
Discovering and exploiting temporal dependence is highly
desired in many applications. For example, consider health care, analysis of temporal dependence can be of value throughout the entire disease treatment process, from disease prevention to treatment. A temporal dependence contains two components: the causative behavior and the dependent behavior. The user’s future behavior is affected by the causative behavior and the causative behavior causes the dependent behavior. For one more example, in E-commerce networks, accurate analysis from users’ time series activity records is of significant importance for advertising, marketing, and psychological studies. Qingchao Ca et al.\cite{DBLP:journals/pvldb/CaiX0JOZ18} proposed  recurrent cohort analysis to group users into cohorts, and Dehua Cheng et al.\cite{DBLP:conf/kdd/ChengB014}  improved the performance of
temporal dependence recovery by  reversing the time stamps of original time series and combine both time series.  
\subsection{Summary and Discussion}

In the field of relational databases, there are many cleaning algorithms based on integrity constraints, which are difficult to apply in the field of time series in which the observed values are substantially numerical, because they follow a strict equality relationship. A few methods, which we summarize  in Table \eqref{Summary of constraints}, can be used for time series data cleaning, for instance ODs and SDs can be used to solve problems in some scenarios, such as the number of miles in a car is non-decreasing. Further speed-based constraints can be used to process data such as GPS and stock prices, but only with relevant domain knowledge can give a reasonable constraint. Therefore, the constraint-based cleaning algorithm needs to be further improved to have better robustness. 
Similarity rules \cite{DBLP:journals/tkde/SongSZCW20}, 
capturing a more general form of constraints on similarities between data values \cite{DBLP:journals/isci/SongZ014} 
for data repairing \cite{DBLP:journals/pvldb/SongCY014,DBLP:journals/vldb/SongLCYC17} 
and imputation \cite{DBLP:journals/pvldb/SongZC015,DBLP:journals/tkde/SongSZCW20}, 
could be considered. 
Moreover, learning individual models \cite{DBLP:conf/icde/ZhangSSW19} could help in repairing missing data in different scenarios. 
One possible future direction is to use anomaly detection methods to detect anomalies first, and then treat outliers as missing to repair. We will discuss anomaly detection in Section V.
\begin{table}
 \centering\small
\caption{Summary of Constraints}
\label{Summary of constraints}
\setlength{\tabcolsep}{3pt}
\begin{tabular}{|p{70pt}|p{150pt}|}
\hline
 Reference& Method\\ \hline
\cite{DBLP:conf/sigmod/DongH82,DBLP:journals/tcs/GinsburgH83,DBLP:conf/pods/GinsburgH83,DBLP:journals/jacm/GinsburgH86} & ODs\\
\cite{DBLP:journals/is/Wijsen98,DBLP:journals/tkde/Wijsen01} & Extend ODs\\
\cite{DBLP:conf/icde/LopatenkoB07} & DCs \\
\cite{DBLP:journals/pvldb/GolabKKSS09} & SDs \\
\cite {beeri1984structure} & FDs\\
\cite{fan2008conditional,DBLP:journals/tkde/FanGLX11} & CFDs \\
\cite{DBLP:conf/sigmod/SongZWY15} & Speed Constraints\\
\cite{DBLP:conf/dasfaa/YinYWHL18} & Variance Constraints \\
\cite{DBLP:journals/tkde/SongSZCW20}& Similarity Rule Constraints \\
\cite{DBLP:conf/icde/ZhangSSW19}& Learning Individual Models \\
\hline
\end{tabular}
\end{table}

\section{Statistics based cleaning algorithm}
\label{sec:Statistics based cleaning algorithm}
Statistical-based cleaning algorithms occupy an important position in the field of data cleaning. Such algorithms use models, which learned from data, to clean data.
The statistical-based approach involves a lot of statistical knowledge, but this article focuses on statistical-based data cleaning methods, so we won't cover statistical-related knowledge in detail.

\subsection{Maximum likelihood}
The intuitive idea of the maximum likelihood principle is a random test, if there are several possible outcomes {$x_1,x_2...x_t$}, if the result $x_i$ occurs in one test, it is generally considered that the test conditions are favorable for $x_i$, or think that $x_i$ has the highest probability of occurrence.

\par{Notation: }For a given time series data $x(t)$, which consistents with a probability distribution $d$, and assume that its probability aggregation function (discrete distribution) is $F_d$; consider a distribution parameter $\theta$,  sampling {$x_1,x_2...x_n$} from this distribution, then use $F_d$ to calculate its probability \cite{DBLP:journals/tsp/BreslerM86} as shown in equation \eqref{maxinum-likelihood}.
\begin{equation}P=(x_1,x_2...x_n)=F_d(x_1,x_2...x_n|\theta)\label{maxinum-likelihood}\end{equation}
\par Ziekow et al. \cite{DBLP:conf/icdt/GogaczT17} use the maximum likelihood technique to clean Radio Frequency Identification (RFID) data. Wang et al. \cite{DBLP:journals/tosn/WangKA14} propose the first maximum likelihood solution to address the challenge of truth discovery from noisy social sensing data. Yakout et al. \cite{DBLP:conf/sigmod/YakoutBE13} argue a new data repairing approach that is based on maximizing the likelihood of replacement data in the given data distribution, which can be modeled using statistical machine learning techniques, but this technology is used to repair the data of the database. For the repairing of time series data errors, Zhang et al. \cite{DBLP:conf/sigmod/ZhangSW16} propose a better solution based on maximum likelihood, which solves the problem from the perspective of probability. According to the probability distribution of the speed change of adjacent data points in the time series, the time series cleaning problem can be converted to find a cleaned time series, which is based on the probability of speed change that has the greatest likelihood.

\subsection{Markov model}
Markov process is a class of stochastic processes, which means that the transition of each state in the process depends only on the previous $n$ states. This process is called a $n-order$ model, where $n$ is the number that affects the transition state. The simplest Markov process is the $first-order$ process, and the transition of each state depends only on the state before it. Time and state are discrete Markov processes called Markov chains, abbreviated as ${X}_{n}=X(n), n=0,1,2...$. The Markov chain \cite{dukhovny1900markov} is a sequence of random variables $X_1, X_2, X_3...$. The range of these variables, that is, the set of all their possible values, is called the "state space", and the value of ${X}_{n}$ is the state of time $n$.

\par The Markov Model \cite{cai1994markov,DBLP:journals/tcom/ZhangK99} is a statistical model based on Markov chain, which is widely used in speech recognition, part-of-speech automatic annotation, phonetic conversion, probabilistic grammar and other natural language processing applications. In order to find patterns that change over time, the Markov model attempts to build a process model that can generate patterns. \cite{cai1994markov} and \cite{DBLP:journals/tcom/ZhangK99} use specific time steps, states, and make Markov assumptions. With these assumptions, this ability to generate a pattern system is a Markov process. A Markov process consists of an initial vector and a state transition matrix. One thing to note about this assumption is that the state transition probability does not change over time.

\par
\begin{figure*}
\includegraphics[width=5 in]{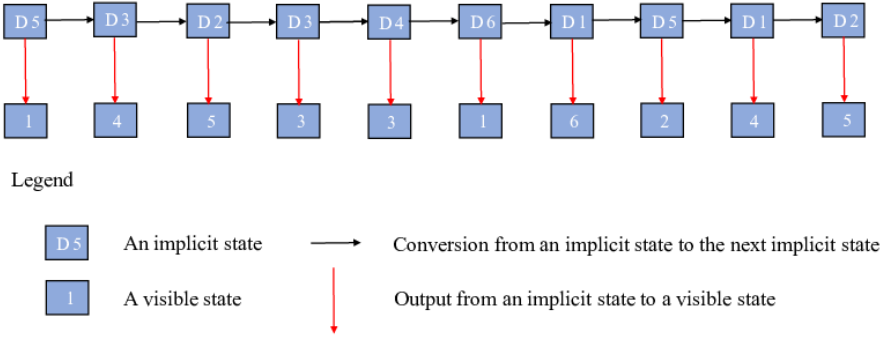}
\caption{An Example of Hidden Markov}
\label{hhm}
\end{figure*}

 Hidden Markov Model (HMM) \cite{DBLP:journals/eswa/HassanNK07,DBLP:journals/eswa/DongYKHEK09} is a statistical model based on Markov Model, which is used to describe a Markov process with implicit unknown parameters. The difficulty is to determine the implicit parameters of the process from observable parameters, and then use these parameters for further analysis, such as prediction of time series data. For instance, after rolling the dice 10 times, we could get a string of numbers, for example we might get such a string of numbers:${1,4,5,3,3,1,6,2,4,5}$ as shown in Figure \ref{hhm}.  This string of numbers is called the visible state chain. But in HMM, we not only have such a string of visible state chains, but also a chain of implied state chains. In this example, the implicit state chain might be: $D5,D3,D2,D3,D4,D6,D1,D5,D1,D2$.

Gupta et al. \cite{gupta2012stock} use HMM to predict the price of stocks. Baba et al. \cite{DBLP:conf/sigmod/BabaJLPKX16} argue a data cleaning method based on the HMM, which used to clean RFID data related to geographic location information. In multi-dimensional time series cleaning, HMM has more application space than the single-dimensional cleaning algorithm, because of the correlation between the dimensions.

\subsection{Binomial Sampling}
Shawn R. Jeffery et al. \cite{DBLP:conf/vldb/JefferyGF06} propose an adaptive method for cleaning RFID data, which exploits techniques based on sampling and smoothing theory to improve the quality of RFID data. Tag transition detection: tag transition detection refers to the fact that when the position of the tag changes, the cleaning result should reflect the fact that the tag leaves. Let's introduce a few RFID-related concepts.
\par(1) Interrogation cycles: the reader's question-and-answer process for the tag is the basic unit of the reader's detection tag.
\par(2) Reader read cycle (epoch, 0.2 sec - 0.25 sec): a collection of multiple interrogation cycles.
\par Based on the above concept, the following definition: $W_i$ is smooth window of tag $i$ and is composed of $\omega_i$ epoch, $S_i$ is the window that tag $i$ is actually detected in the $W_i$ window, $Count_t$ indicates the number of inquiry cycles of $t$, and $R$ is the corresponding number of $t$ epoch tag $i$. For a given time window, suppose the probability that the tag $i$ may be read in each epoch is $p_i=\frac{R}{Count_t}$, and the Statistical Soothing for Unreliable RFID Data (SMURF) \cite{DBLP:conf/vldb/JefferyGF06} algorithm treats each epoch's reading of the tag as a Bernoulli experiment with probability $p_i$. Therefore, $p_i$ conforms to the binomial distribution $B(\omega_i,p_i)$. ${p}_{i,avg}$ is the average read rate in $S_i$.
\par Using the model based on the Bernoulli experiment to observe the tag $i$, if the average reading rate of the tags in $\omega_i$ epoch is $(1-p_{i,avg})^{\omega_i}$.
 To ensure the dynamic nature of the tag the size of the sliding window $W_i$ needs to be satisfied as shown in equation \eqref{smurf}.
\begin{equation}||S_i|-\omega_ip_{i,avg}|>2\sqrt{\omega_ip_{i,avg}(1-p_{i,avg})}\label{smurf}\end{equation}
\par
The SMURF algorithm first sets the initial window size to 1, and then dynamically adjusts the window length based on the actual situation of the read. If the current window meets the integrity requirement \cite{DBLP:conf/vldb/JefferyGF06}, the SMURF algorithm will detect the status of the tag. When the detection result indicates that the tag status changes, SMURF will adjust the current window length to 1/2 of the original window to react to the tag's transition. If the calculated window size that satisfies the integrity constraint is greater than the current window size, the algorithm linearly increases the current window size by 2 steps and outputs the point data in the current window. If it is detected that the label does not move, the algorithm outputs the current window midpoint as the output point, and then continues to slide an epoch for the next processing. 
\par SMURF algorithm is widely used to clean RFID data, and many studies \cite{DBLP:journals/ijguc/XuDLSW18,leema2011effective} improve it. Leema, A et al. \cite{leema2011effective} study the effect of tag movement speed on data removal results and H Xu et al. \cite{DBLP:journals/ijguc/XuDLSW18} consider the impact of data redundancy on setting up sliding windows.

\subsection{Spatio-Temporal Probabilistic Model}
Besides data cleaning, M.~Milani et al. \cite{DBLP:conf/icde/ZhengMC19} propose Spatio-Temporal Probabilistic Model (STPM), this method learns more detailed data patterns from historical data, and then cleans the current data. STPM not only gives joint probability distributions that are updated on the data set at different times, but also distinguishes association updates from association values. STPM based on Dynamic Probabilistic Relational Models (DRPMs), so we need to state DRPMs model first. The DRPMs is a graph model used to represent the relationship between dynamic data sets, its models based on the dependency relationship between attributes, and generally uses conditional probability distribution to calculate the probability of each attribute value in a given parent node value and forms a relationship chain. For instance, when we need to estimate the data at time $T$, we can only use the data before time $T$ to infer, namely, the current state depends only on the previous state, which is similar to the Markov Model. STPM extends DRPMs to model update pattern between different time data, and captures spatial and temporal update patterns by modeling updates events to provide update relationships of possible existence, finally detect and repair data.

\subsection{Others}
Firstly, we summarize the methods described above in Table \eqref{Summary of statistics}. In fact, Bayesian prediction model is a technique based on Bayesian statistics. The Bayesian prediction model utilizes model information, data information, and prior information, so the prediction effect is good, there this model is widely used, including in the field of time series data cleaning. Wang et al. \cite{DBLP:conf/sigmod/WangDM15} establish a cost model for Bayesian analysis which is used to analyze errors in the data. Bergman et al. \cite{DBLP:conf/sigmod/BergmanMNT15} consider the user's participation and use the user's feedback on the query results to clean the data. Mayfield et al. \cite{DBLP:conf/sigmod/MayfieldNP10} propose a more complex relationship-dependent network (RDN \cite{DBLP:journals/jmlr/NevilleJ07}) model to model the probability relationships between attributes. The difference between RDN and traditional relational dependencies (such as Bayesian networks \cite{DBLP:conf/icml/GetoorFKT01}) is that RDNs can contain ring structures. The method iteratively cleans the data set and observes the change in the probability distribution of the data set before and after each wash. When the probability distribution of the data set converges, the cleaning process is aborted. Zhou et al. \cite{DBLP:conf/sigmod/ZhouT15} argue a technique for accelerating the learning of Gaussian models via using GPU. The article believes that in the case of excessive data, it is not necessary to use all the data to learn the model. Also, the author provides a method of automatic tuning. In order to clean and repair fuel level data, Tian et al. \cite{DBLP:journals/tii/TianZDHSCWW19}  propose a modified Gaussian mixture model (GMM) based on the synchronous iteration method, which uses the particle swarm optimization algorithm and the steepest descent algorithm to optimize the parameters of GMM and uses linear interpolation-based algorithm to correct data errors. Shumway et al. \cite{shumway1982approach} use the EM \cite{dempster1977maximum} algorithm combined with the spatial state model \cite{jones1966exponential,morrison1977kalman} to predict and smooth the time series.

\begin{table}
 \centering\small
\caption{Summary of Statistics}
\label{Summary of statistics}
\setlength{\tabcolsep}{3pt}
\begin{tabular}{|p{70pt}|p{150pt}|}
\hline
 Reference& Method\\ \hline
\cite{DBLP:journals/tsp/BreslerM86,DBLP:conf/icdt/GogaczT17,DBLP:journals/tosn/WangKA14,DBLP:conf/icde/AggarwalY15} 
\par \cite{DBLP:conf/sigmod/YakoutBE13,DBLP:conf/sigmod/ZhangSW16}& Maximum Likelihood\\
\cite{dukhovny1900markov,cai1994markov,DBLP:journals/tcom/ZhangK99} & Markov Model\\
\cite{DBLP:journals/eswa/HassanNK07,DBLP:journals/eswa/DongYKHEK09,gupta2012stock}
\par \cite{DBLP:conf/sigmod/BabaJLPKX16}& HMM\\
\cite{leema2011effective,DBLP:journals/ijguc/XuDLSW18,DBLP:conf/vldb/JefferyGF06}& SMURF\\
\cite{DBLP:conf/icde/ZhengMC19}& STPM\\
\cite{DBLP:conf/sigmod/WangDM15,DBLP:conf/icml/GetoorFKT01}& Bayesian\\
\cite{DBLP:conf/sigmod/BergmanMNT15,DBLP:journals/jmlr/NevilleJ07,DBLP:conf/sigmod/MayfieldNP10} & RDN \\
\cite{DBLP:journals/tii/TianZDHSCWW19} &GMM \\
\cite{shumway1982approach} &EM\\
\hline
\end{tabular}
\end{table}

\section{Time series anomaly detection}
\label{sec:Time series anomaly detection}
Gupta et al. \cite{DBLP:journals/tkde/GuptaGAH14} investigate the anomaly detection methods for time series data: for a given time series data, there may be two types of outliers, namely single-point anomalies and subsequence anomalies (continuous anomalies). In this section, we first discuss the detection methods of abnormal points and abnormal sequences, next introduce the application of Density-Based Spatial Clustering of Applications with Noise (DBSCAN) algorithm in data cleaning, and then review the abnormal detection methods related to machine learning.

\subsection{Abnormal point detection}
\begin{figure}
\includegraphics[width=3 in]{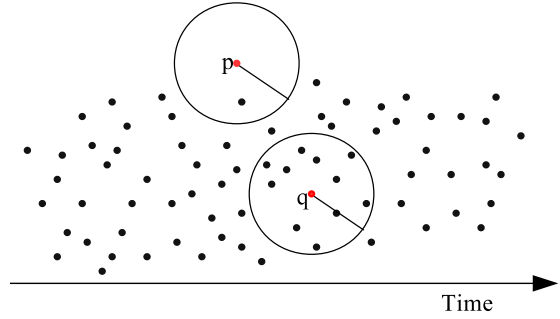}
\caption{An Example Abnormal Point Detection}
\label{outlier_point}
\end{figure}

 For single-point anomalies, the most common idea is to use predictive models for detection. That is, the predicted value of the established model and the observed value for each data point is compared, and if the difference between the two values is greater than a certain threshold, the observed value is considered to be an abnormal value. Specifically, Basu et al. \cite{DBLP:journals/kais/BasuM07} select all data points with timestamps $t-k$ to $t+k$ with the timestamp $t$ as the center point, and the median of these data points is considered to be the predicted value of data points with timestamp $t$ value. Hill et al. \cite{DBLP:journals/envsoft/HillM10} first cluster the data points and take the average of the clusters as the predicted value of the point. The AR model and the ARX model are widely used for anomaly detection in various fields, such as economics, social surveys \cite{brockwell2016introduction,box2015time}, and so on. The ARX model takes advantage of manually labeled information, so it is more accurate than the AR model when cleaning data. The ARIMA model \cite{DBLP:journals/ijon/Zhang03} represents a type of time series model consisting of AR and MA mentioned above, which can be used for data cleaning of non-stationary time series.
Kontaki et al. \cite {DBLP:conf/icde/KontakiGPTM11} propose continuous monitoring of distance-based outliers over data streams. One of the most widely used definitions is the one based on distance as shown in Figure \ref{outlier_point}: an object $p$ is marked as an outlier, if there are less than $k$ objects in given distance. Here $k=4$, $q$ is the normal point and $p$ is the abnormal point.

\subsection{Abnormal sequence detection}
\begin{figure}
\includegraphics[width=3 in]{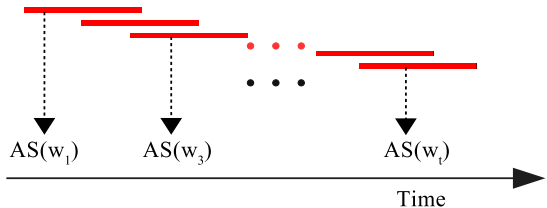}
\caption{An Example Abnormal Sequence Detection}
\label{slide_window}
\end{figure}

Different studies have different definitions of subsequence anomalies. Keogh et al. \cite{DBLP:journals/kais/KeoghLLH07} proposed that a subsequence anomaly, that is, a subsequence has the largest distance from its nearest non-overlapping match. With this definition, the simplest calculation method is to calculate the distance between each subsequence with length $n$ and other subsequences. Of course, the time complexity of this calculation method is very high. In the later studies, Keogh et al. \cite{DBLP:conf/icdm/KeoghLF05} propose a heuristic algorithm by reordering candidate subsequences and Wei et al. \cite{DBLP:conf/icdm/WeiKX06} argue an acceleration algorithm using local sensitive hash values. In calculating the distance, the Euclidean distance is usually used, and Keogh \cite{DBLP:conf/kdd/KeoghLR04} further proposes a method using the compression-based similarity measure as the distance function. As shown in Figure \ref{slide_window}, the data is divided into multiple sub-sequences that overlap each other. First, calculate the abnormal score of each window, and then calculate the abnormal score (AS) of the whole test sequence according to the abnormal score (AS) of each window. Window-based techniques can better locate anomalies compared to direct output of the entire time series as outliers. There are two main types of methods based on this technique. One is to maintain a normal database \cite {DBLP:conf/id/GhoshSS99,DBLP:conf/acsac/Endler98}, and then compare the test sequence with the sequence in the normal database to determine whether it is abnormal; the other is to build an anomalous database \cite {DBLP:journals/gpem/GonzalezD03,dasgupta2002anomaly} and then compare the test sequence with the sequence in the database to detect if it is anomalous. \cite{DBLP:conf/edbt/Senin0WOGBCF15} found that the length of the error is unknown and they use grammar induction to aid anomaly detection without any prior knowledge.

\subsection{Density-Based Spatial Clustering of Applications with Noise}
The DBSCAN \cite{DBLP:conf/kdd/EsterKSX96} algorithm is a clustering method based on density-reachable relationship, which divides the region with sufficient density into clusters and finds clusters of arbitrary shape in the spatial database with noise and defines the cluster as the largest set of points connected by density. Then the algorithm defines the cluster according to the set density threshold as the basis for dividing the cluster, that is, when the threshold is satisfied, it can be considered as a cluster.

The principle of DBSCAN algorithm:
(1) DBSCAN searches for clusters by checking the $Eps$ neighborhood of each point in the data set. If the $Eps$ neighborhood of point $p$ contains more points than $MinPts$, create a cluster with $p$ as the core object;
(2) Then, DBSCAN iteratively aggregates objects that are directly reachable from these core objects. This process may involve the consolidation of some density-reachable clusters;
(3) When no new points are added to any cluster, the process ends.
\par
\begin{figure}
\hfill
\includegraphics[width=3in]{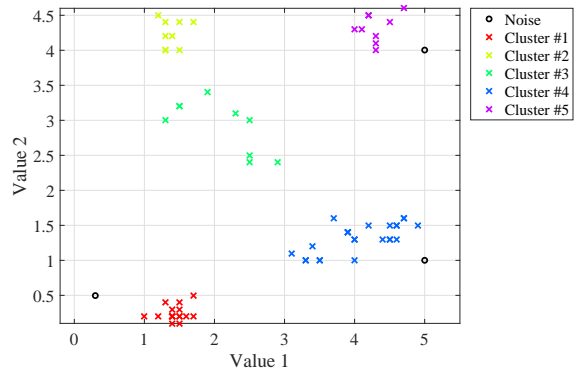}
\caption{An Example of DBSCAN Clustering}
\label{dbscan}
\end{figure}

 Where $MinPts$ is the minimum number of neighbor points that a given point becomes the core object in the neighborhood, $Eps$ is the neighborhood radius. For instance, 
$Eps$ is 0.5 and $MinPts$ is 3, for a given data set, the effect of clustering is as shown in Figure \ref{dbscan}. Some noise points can be repaired and clustered into classes adjacent to them. Recent research \cite{DBLP:conf/kdd/SongLZ15} has shown that after repairing erroneous data. They also perform cleaning experiments on GPS data based on DBSCAN, the accuracy of clustering on spatial data can be improved. But this method cannot solve continuous errors and needs further improvement. 

\subsection{Speed-based cleaning algorithm }
\begin{figure}
\hfill
\includegraphics[width=3.5in]{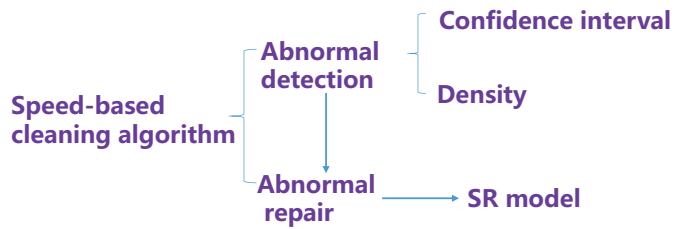}
\caption{ The Architecture Diagram of SC Algorithm}
\label{sc}
\end{figure}

The speed-based cleaning algorithm\cite{wangxi} (SC algorithm) performs anomaly detection through the confidence interval in the time dimension and density in the spatial dimension as shown in Figure \ref{sc}. And designs an algorithm for detecting continuous errors based on the above two detection methods. Wang et al.\cite{wangxi} propose the SR model for repairing the data detected as abnormal. SC algorithm is mainly used for cleaning irregular time time series data. To solve the problem of irregular time series data cleaning, they use the speed characteristics of irregular time series data to design SC algorithms that can be used to clean rregular time series data. Wang et al.\cite{wangxi} prove that the SR model is equivalent to the AR model at equal interval time series, and gives an incremental speed-based cleaning algorithm (ISC algorithm). The time complexity of the SC algorithm is relatively high, and pruning can be considered in the future to reduce the time complexity.

\subsection{Generative Adversarial Networks}
With the rapid development of machine learning technology, more and more problems are solved using machine learning. Dan Li et al. \cite{DBLP:conf/icann/LiCJSGN19} use the GANs network to effectively detect anomalies in time series data.
GANs trains two models at the same time, which are the generation model for capturing data distribution and the discriminant model for discriminating whether the data are real data or pseudo data as shown in Figure \ref{Generative Adversarial Networks}. 

Given a random variable with a probability of uniform distribution as input, we want to generate a probability distribution of the output as "dog probability distribution". The philosophy of Generative Matching Networks (GMNs), which idea is to train the generative network by directly comparing the generated distribution with the true distribution, is to optimize the network by repeating the following steps:
\par (1) Generate some evenly distributed input;
\par (2) Let these inputs go through the network and collect the generated output;
\par (3) Compare the true "dog probability distribution" with the generated "dog probability distribution" based on the available samples (e.g. calculate the MMD distance between the real dog image sample and the generated image sample);
\par (4) Use backpropagation and gradient descent to compute the errors and update the weights. The purpose of this process is to minimize the loss of the generation model and discriminant.
\par Dan Li et al. \cite{DBLP:conf/icann/LiCJSGN19} use GANs to detect abnormalities in time series and a natural idea is to use GANs network to repair missing values of time series data. Perhaps more machine learning algorithms are waiting for the cleaning of time series error values. A simple idea is to treat the detected anomaly data as missing data and then repair it. {Y.~Sun et al. \cite{DBLP:conf/itsc/SunPLS18} first analyze the similarity between parking space data and parking data, and then use Recurrent GANs to generate parking data as repair data, which provide a new idea for solving the problem of time series data repair. C.~Fang et al. \cite{DBLP:conf/cikm/FangSCG19} propose FuelNet which is based on Convolutional Neural Networks (CNNs) and GANs. FuelNet is used to repair the inconsistent and impute the incomplete fuel consumption rates over time.

\begin{figure}
\includegraphics[width=3 in]{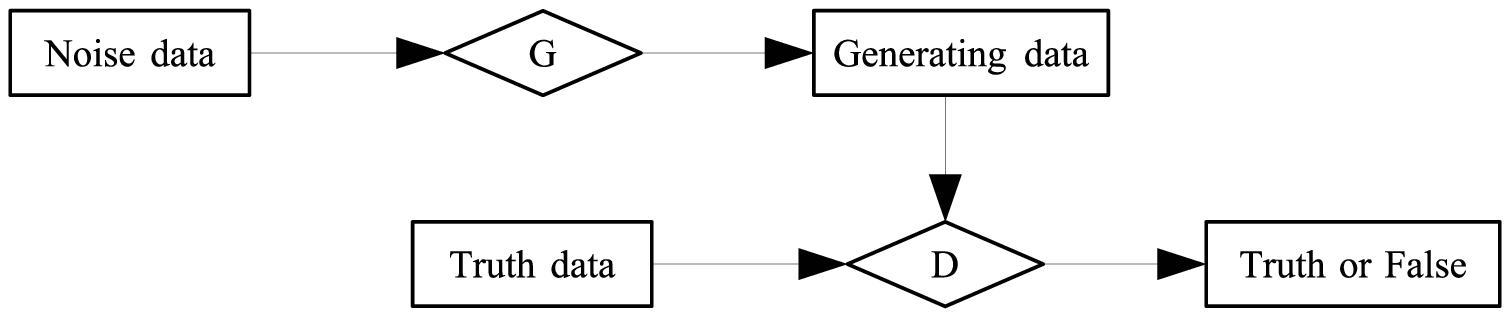}
\caption{A Simple Flow Chart of Generative Adversarial Networks}
\label{Generative Adversarial Networks}
\end{figure}

\subsection{Long Short-Term Memory}
Since Recurrent Neural Network (RNN) also has the problem of gradient disappearance, it is difficult to process long-sequence data. F. A. Gers et al. \cite {DBLP:journals/neco/GersSC00} improve RNN and got the RNN special case Long Short-Term Memory (LSTM), which can avoid the disappearance of the regular RNN gradient. It has been widely used in industry and 
\cite{DBLP:journals/corr/FilonovLV16,DBLP:journals/corr/MalhotraTRAVAS16,DBLP:conf/esann/MalhotraVSA15} use LSTM to perform anomaly detection on time series data.

\begin{figure}
\includegraphics[width=3 in]{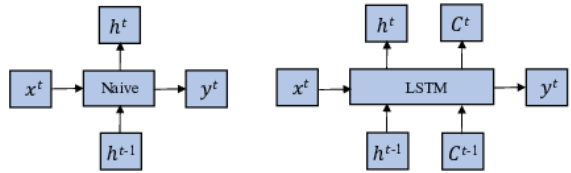}
\caption{Simple RNN Structure and  Simple LSTM Structure}
\label{lstm}
\end{figure}
The left picture is a simple RNN structure diagram, and the right picture is a simple LSTM structure diagram in Figure \ref{lstm}, where given function as shown in equation \eqref{rnn}.
\begin{equation}F: h,y=f(h,x)\label{rnn}\end{equation}
\par
In equation \eqref{rnn}, ${x}^{t}$ is the input of data in the current state, ${h}^{t-1}$ (hidden state) indicates the input of the previous node received, ${y}^{t}$ is the output in the current state and ${h}^{t}$ is the output passed to the next node. As can be seen from the Figure \ref{lstm}, the output ${h}^{t}$ is related to the values of ${x}^{t}$ and ${h}^{t-1}$. ${y}^{t}$ is often used to invest in a linear layer (mainly for dimension mapping), and then use $softmax$ to classify the required data. As shown in Figure \ref{lstm}, RNN has only one delivery state ${h}^{t}$, LSTM also has a delivery status ${c}^{t}$ (cell state). There are three main stages within LSTM:
\par (1) Forgotten phase. The forgetting phase is mainly to forget the input that is passed in from the previous node. A simple principle is: forget the unimportant, remember the important one. More specifically,  ${z}^{f}$ is calculated as a forgotten gate, which is used to control the previous state ${c}^{t-1}$ , and then decide whether to retain the data or forget it.
\par (2) Selective memory phase. At this stage, the input is selectively memorized. Mainly to remember the input $x$, the more important the data needs to be more reserved.
\par (3) Output phase. This phase determines which outputs would be treated as current states. Similar to the normal RNN, the output ${y}^{t}$ is often also obtained by ${h}^{t}$ change.

\par Filonov, Pavel et al. \cite{DBLP:journals/corr/FilonovLV16} and Pankaj Malhotra et al. \cite{DBLP:journals/corr/MalhotraRAVAS16} provide recurrent neural networks by providing network time series data. The recurrent neural network understands what the normal expected network activity is. When an unfamiliar activity from the network is provided to a trained network, it can distinguish whether the activity is expected or invaded.

\subsection{Summary and Discussion}

In addition to the methods described above, we also summarize some common methods in Table \eqref{Summary of detection}. As shown in Table \eqref{Summary of detection}, Xing et al. \cite {DBLP:journals/kais/XingPY12} show that the cleaned sequence can improve the accuracy of time series classification. 
Diao et al. \cite {DBLP:conf/cyberc/DiaoLMYH15} design LOF \cite {DBLP:conf/sigmod/BreunigKNS00} based online anomaly detection and cleaning algorithm.
Zhang et al. \cite {DBLP:journals/pvldb/ZhangS0Y17}  propose an iterative minimum cleaning algorithm based on the timing correlation of error time series in continuous errors and keep the principle of minimum modification in data cleaning. The algorithm is effective in cleaning continuous errors in time series data. 
Qu et al. \cite {qu2016data} first use cluster-based methods for anomaly detection and then use exponentially weighted averaging for data repair, which is used to clean power data in a distributed environment. R.~Corizzo et al. \cite{DBLP:journals/bdr/CorizzoCJ19} use detect anomalous geographic data by distance-based method, and then use Gradient-boosted tree (GBT) to repair the anomalous data. Charu C et al.\cite{DBLP:conf/icde/AggarwalY15} proposed a solution for distributed storage and query for large-scale streaming sensor data, and they examined the problem of historical storage and diagnosis of massive numbers of simultaneous
streams. We can conclude that anomaly detection algorithms play an important role in time series data cleaning. It is also becoming more and more important to design anomaly detection algorithms for time series repair, and we discuss future directions in Section \ref{sec:Conclusion and Future Directions}.

\begin{table}
 \centering\small
\caption{Summary of Detection}
\label{Summary of detection}
\setlength{\tabcolsep}{3pt}
\begin{tabular}{|p{70pt}|p{150pt}|}
\hline
 Reference& Method\\ \hline
\cite{DBLP:journals/kais/BasuM07,DBLP:journals/kais/KeoghLLH07,DBLP:conf/icdm/KeoghLF05} \par
\cite{DBLP:conf/cyberc/DiaoLMYH15,DBLP:journals/bdr/CorizzoCJ19} & Distance-based\\
\cite{wangxi} &SC \\
\cite{DBLP:conf/id/GhoshSS99,DBLP:conf/acsac/Endler98} & Maintain a Normal Database\\
\cite{DBLP:journals/gpem/GonzalezD03,dasgupta2002anomaly} & Build an Anomalous Database\\
\cite{DBLP:conf/kdd/EsterKSX96,DBLP:conf/kdd/SongLZ15,qu2016data} & Clustering\\
\cite{DBLP:conf/icann/LiCJSGN19,DBLP:conf/itsc/SunPLS18,DBLP:conf/cikm/FangSCG19} & GANs\\
\cite{DBLP:journals/corr/FilonovLV16,DBLP:journals/corr/MalhotraTRAVAS16,DBLP:conf/esann/MalhotraVSA15} & LSTM \\
\hline
\end{tabular}
\end{table}

\section{Tools and Evaluation criteria}
\label{sec:Tools and Evaluation criteria}
In this section, we first give an overview of tools to clean time series and then summarize evaluation criteria related to time series cleaning methods.
\subsection{Tools}

\begin{table*}
 \centering\small
\caption{Some Examples of Tools or Systems}
\label{tools}
\setlength{\tabcolsep}{3pt}
\begin{tabular}{|p{70pt}|p{90pt}|p{310pt}|}
\hline
 &Method& Detail\\\hline
PIClean \cite{DBLP:conf/sigmod/YuC19}&  Based on statistics & Produce probabilistic errors and probabilistic fixes using low-rank approximation, which implicitly discovers and uses relationships between columns of a dataset for cleaning. \\

HoloClean \cite{DBLP:journals/pvldb/RekatsinasCIR17}&  Based on statistics & Learn the probability model and then select the final data cleaning plan based on the probability distribution. \\

ActiveClean \cite {DBLP:journals/pvldb/KrishnanWWFG16}&  Based on statistics & Allow for progressive and iterative cleaning in
statistical modeling problems while preserving convergence guarantees. \\

Cleanits \cite {DBLP:journals/pvldb/DingWSLLG19}&  Anomaly detection & Develop reliable data cleaning algorithms by considering features of both industrial time series and domain knowledge. \\

MLClean \cite {DBLP:conf/sigmod/TaeROKW19}&  Anomaly detection & The combination of data cleaning technology and machine learning methods is designed to generate unbiased cleaning data, which is used to train accurate models. \\

ASAP  \cite {DBLP:journals/pvldb/RongB17}&  Smoothing based & Develop a new analytics operator called ASAP that automatically smooths streaming time series by adaptively optimizing the trade-off between noise reduction and trend retention. \\

EDCleaner \cite {DBLP:conf/icc/0004ZFWYY19} &  Based on statistics & For social network data, detection and cleaning are performed through the characteristics of statistical data fields. \\

PACAS \cite {DBLP:conf/bigdataconf/HuangMC18}& Based on statistics  &Design a framework for data cleaning between service providers and customers.\\

TsOutlier \cite {DBLP:conf/bigdataconf/HuangCLS019}& Anomaly detection  &Use multiple algorithms to detect anomalies in time series data, and support both batch and streaming processing.\\
EGADS \cite{DBLP:conf/kdd/LaptevAF15}& Anomaly detection & Offer two classes of algorithms for detecting outliers: Plug-in methods and Decomposition-based methods, which is designed for automatic anomaly detection of large-scale time series data.\\
\hline
\end{tabular}
\end{table*}

 There are many tools or systems for data cleaning, but they are not effective on time series cleaning problems. In Table \ref{tools} we investigate some tools that might be used for time series cleaning because they \cite {DBLP:journals/pvldb/RekatsinasCIR17,DBLP:conf/sigmod/YuC19,DBLP:conf/sigmod/TaeROKW19,DBLP:journals/pvldb/KrishnanWWFG16} {are originally used to solve traditional database cleaning problems. Ding et al. \cite {DBLP:journals/pvldb/DingWSLLG19} present Cleanits, which is an industrial time series cleaning system and implements an integrated cleaning strategy for detecting and repairing in industrial time series. Cleanits provides a user-friendly interface so users can use results and logging visualization over every cleaning process. Besides, the algorithm design of Cleanits also considers the characteristics of industrial time series and domain knowledge. The ASPA proposed by Rong et al. \cite {DBLP:journals/pvldb/RongB17} violates the principle of minimum modification and distort the data, which is not suitable for being used widely. 
EDCleaner proposed by J.~Wang et al. \cite{DBLP:conf/icc/0004ZFWYY19} is designed for social network data, detection and cleaning are performed through the characteristics of statistical data fields.
 Y.~Huang et al.\cite{DBLP:conf/bigdataconf/HuangMC18} propose PACAS which is a framework for data cleaning between service providers and customers.
R.~Huang et al. \cite{DBLP:conf/bigdataconf/HuangCLS019} present TsOutlier, a new framework for detecting outliers with explanations over IoT data. TsOutlier uses multiple algorithms to detect anomalies in time series data, and supports both batch and streaming processing.
 EGADS \cite{DBLP:conf/kdd/LaptevAF15} Offers two classes of algorithms for detecting outliers: Plug-in methods and Decomposition-based methods, which is designed for automatic anomaly detection of large-scale time series data.
There is not much research on time series cleaning tools or systems, and we discuss further in Future Directions in Section~\ref{sec:Conclusion and Future Directions}.
\subsection{Evaluation criteria}
The Root Mean Square (RMS) error \cite{DBLP:conf/vldb/JefferyGF06} is used to evaluate the effectiveness of the cleaning algorithm. Let $x$ denotes the sequence consisting of the true values of the time series, $\overline{x}$ denotes the sequence consisting of the observations after the error is added, and $\widehat{x}$ denotes the sequence consisting of the repaired values after the cleaning. Here the RMS error \cite {DBLP:conf/vldb/JefferyGF06} is represented as shown in equation \eqref{rms}.
\begin{equation}\Delta(x,\widehat{x})=\sqrt{\frac{1}{n}\sum^n_{i=1}(x_i-\widehat{x}_i)^2}\label{rms}\end{equation}
\par The equation \eqref{rms}  measures the distance between the true value and the cleaned value. The smaller the RMS error, the better the cleaning effect.
\par Other criteria include error distance between incorrect data and correct data, repaired distance between erroneous data and cleaned results (as shown in equation \eqref{repair-distance} referring to the minimum modification principle in data repairing).
\begin{equation}\Delta(\overline{x},\widehat{x})=\sum^n_{i=1}|\overline{x_i}-\widehat{x}_i|\label{repair-distance}\end{equation}
\par Dasu et al. \cite {DBLP:journals/pvldb/DasuL12} propose a statistical distortion method to evaluate the quality of cleaning methods. The proposed method directly observes the numerical distribution in the data set and evaluates the quality according to the variation of the distribution caused by different cleaning methods. 

\section{Conclusion and Future Directions}
\label{sec:Conclusion and Future Directions}
In this paper, we review four types of time series cleaning algorithms, cleaning tools or systems and related research on evaluation criteria. Next, we summarize the full text in Section \ref{sec:Conclusion} and list some advice of future directions in Section \ref{sec:Future Directions}.
\subsection{Conclusion}
\label{sec:Conclusion}
With the development of technology, people gradually realize the value contained in the data.
Owing to companies want to derive valuable knowledge from these data, and data analysis has played an increasingly important role in finance, healthcare, natural sciences, and industry. Time series data, as an important data type, is widely found in industrial manufacturing. For instance, a wind power enterprise analyzes sensor data, which are located throughout the wind turbine, to determine whether the fan is in a normal state; transport companies also want to optimize vehicle fleet travel by analyzing vehicle GPS information. However, due to external environmental interference, sensor accuracy, and other issues, time series data often contain many errors that can interfere with subsequent data analysis and cause unpredictable effects.

\subsection{Future Directions}
\label{sec:Future Directions}
As mentioned above, data is an intangible asset and advanced technology helps to fully exploit the potential value of data. Thereby, time series data cleaning methods provide very important technical support for the discovery of these values in processing time series error data. Next we list some advice of future directions based on \cite{zhangaoqian}. 

\par The error type illustrates handbook of time series data.
At present, data scientists have a very detailed analysis of the errors in the traditional relational database. However, there is still much work to be further studied in the analysis of time series data error types. For instance, this paper roughly divides the types of time series errors into three types, namely single point big errors, single point small errors and continuous errors. In fact, in continuous errors, there are also a lot of meticulous types of errors, such as additive errors, innovational errors \cite{tsay1988outliers} 
or missing errors \cite{DBLP:journals/pvldb/0001SZL13,DBLP:journals/tkde/0001SZLS16}. How to systematically analyze these error types and form time series data error type illustrated handbook is very important. The clear error type helps to develop targeted cleaning algorithms to solve the problem of ``GIGO (Garbage in, garbage out.)'' that exists in the current field.
	
\par The design of time series data cleaning algorithm. Each chapter of this paper reviews some time series error cleaning algorithms, but further optimizations are possible. The existing methods are mostly for a single-dimensional time series (even the GPS data exists two dimensions' information), but each dimension is cleaned separately during cleaning \cite{DBLP:journals/pvldb/ZhangS0Y17,DBLP:conf/sigmod/SongZWY15,DBLP:conf/sigmod/ZhangSW16}. To further improve the practicability of the algorithm, it is imperative to consider the cleaning of multidimensional time series. Besides, with the development of machine learning technology, more technical learning techniques should be considered for data cleaning algorithms, which may lead to better cleaning results because of the mathematical support behind them.

\par The implementation of time series cleaning tool.  At present, the mainstream data cleaning tools in the industry are still aimed at relational databases, and these tools are not ideal for processing time series data. As time series data cleaning problems become more serious, how to use the fast-developing distributed technology, high performance computing technology and stream processing technology to implement time series cleaning tools (including research tools and commercial tools) and apply them to real-world scenarios such as industry is also the key work of the next stage.  

\par The algorithm design of time series anomaly detection. In real-world scenarios, efficient anomaly detection algorithms play an irreplaceable role in time series repair. It is difficult to judge the difference between the error value and the true value, so it is necessary to specifically design a time series anomaly detection algorithm that can be applied to an industrial scene. It is worth noting that more research is needed on how to perform anomaly detection, cleaning, and analysis in the case of weak domain knowledge or less labeled data.

\par Design of data cleaning algorithms for specific application scenarios. With the application of various technologies in the industry, the application scenarios are becoming more and more clear. The requirements for cleaning algorithms in different application scenarios have different focuses. 
For instance, rather than cleaning errors, another application is to directly answer queries over the possible repairs \cite{DBLP:conf/sigmod/LianCS10}. 
Moreover, rather than quantitive values, cleaning qualitative events \cite{DBLP:conf/sigmod/ZhuSL0Z14,DBLP:journals/tkde/GaoSZWLZ18} 
under pattern constraints \cite{DBLP:conf/icde/ZhuSWYS14,DBLP:journals/tkde/SongGWZWY17}
is also highly demanded.
Finally, the data stored in the Blockchain network \cite {DBLP:conf/sensys/Casado-VaraPPC18} are generally structured data, with the development of Blockchain technology, the design of data cleaning algorithms on Blockchain networks is also particularly important.

\bibliographystyle{unsrt}
\bibliography{timeseries}

\begin{thebibliography}{100}

\bibitem{shumway2017time}
Robert~H Shumway and David~S Stoffer.
\newblock {\em Time series analysis and its applications: with R examples}.
\newblock Springer, 2017.

\bibitem{hamilton1994time}
James~Douglas Hamilton.
\newblock {\em Time series analysis}, volume~2.
\newblock Princeton university press Princeton, NJ, 1994.

\bibitem{brockwell2016introduction}
Peter~J Brockwell and Richard~A Davis.
\newblock {\em Introduction to time series and forecasting}.
\newblock springer, 2016.

\bibitem{box2015time}
George~EP Box, Gwilym~M Jenkins, Gregory~C Reinsel, and Greta~M Ljung.
\newblock {\em Time series analysis: forecasting and control}.
\newblock John Wiley \& Sons, 2015.

\bibitem{DBLP:conf/pervasive/JefferyAFHW06}
Shawn~R. Jeffery, Gustavo Alonso, Michael~J. Franklin, Wei Hong, and Jennifer
  Widom.
\newblock Declarative support for sensor data cleaning.
\newblock In {\em Pervasive Computing, 4th International Conference,
  {PERVASIVE} 2006, Dublin, Ireland, May 7-10, 2006, Proceedings}, pages
  83--100, 2006.

\bibitem{DBLP:journals/debu/DasuDS16}
Tamraparni Dasu, Rong Duan, and Divesh Srivastava.
\newblock Data quality for temporal streams.
\newblock {\em {IEEE} Data Eng. Bull.}, 39(2):78--92, 2016.

\bibitem{shilakes1998enterprise}
C~Shilakes and J~Tylman.
\newblock Enterprise information portals. enterprise software team.
\newblock {\em Enterprise Information Portals}, 1998.

\bibitem{DBLP:journals/pvldb/SongC016}
Shaoxu Song, Yue Cao, and Jianmin Wang.
\newblock Cleaning timestamps with temporal constraints.
\newblock {\em {PVLDB}}, 9(10):708--719, 2016.

\bibitem{DBLP:conf/sigmod/ChuIKW16}
Xu~Chu, Ihab~F. Ilyas, Sanjay Krishnan, and Jiannan Wang.
\newblock Data cleaning: Overview and emerging challenges.
\newblock In {\em Proceedings of the 2016 International Conference on
  Management of Data, {SIGMOD} Conference 2016, San Francisco, CA, USA, June 26
  - July 01, 2016}, pages 2201--2206, 2016.

\bibitem{hellerstein2008quantitative}
Joseph~M Hellerstein.
\newblock Quantitative data cleaning for large databases.
\newblock {\em United Nations Economic Commission for Europe (UNECE)}, 2008.

\bibitem{DBLP:journals/pvldb/KhayatiLTC20}
Mourad Khayati, Alberto Lerner, Zakhar Tymchenko, and Philippe
  Cudr{\'{e}}{-}Mauroux.
\newblock Mind the gap: An experimental evaluation of imputation of missing
  values techniques in time series.
\newblock {\em {PVLDB}}, 13(5):768--782, 2020.

\bibitem{DBLP:journals/jnca/KarkouchMMN16}
Aimad Karkouch, Hajar Mousannif, Hassan~Al Moatassime, and Thomas No{\"{e}}l.
\newblock Data quality in internet of things: {A} state-of-the-art survey.
\newblock {\em J. Network and Computer Applications}, 73:57--81, 2016.

\bibitem{tsay1988outliers}
Ruey~S Tsay.
\newblock Outliers, level shifts, and variance changes in time series.
\newblock {\em Journal of forecasting}, 7(1):1--20, 1988.

\bibitem{DBLP:journals/tkde/GuptaGAH14}
Manish Gupta, Jing Gao, Charu~C. Aggarwal, and Jiawei Han.
\newblock Outlier detection for temporal data: {A} survey.
\newblock {\em {IEEE} Trans. Knowl. Data Eng.}, 26(9):2250--2267, 2014.

\bibitem{zhangaoqian}
Aoqian Zhang.
\newblock {\em Research on Time Series Data Cleaning}.
\newblock PhD thesis, Tsinghua University, 2018.

\bibitem{DBLP:conf/sigmod/BohannonFFR05}
Philip Bohannon, Michael Flaster, Wenfei Fan, and Rajeev Rastogi.
\newblock A cost-based model and effective heuristic for repairing constraints
  by value modification.
\newblock In {\em Proceedings of the {ACM} {SIGMOD} International Conference on
  Management of Data, Baltimore, Maryland, USA, June 14-16, 2005}, pages
  143--154, 2005.

\bibitem{DBLP:conf/icdt/AfratiK09}
Foto~N. Afrati and Phokion~G. Kolaitis.
\newblock Repair checking in inconsistent databases: algorithms and complexity.
\newblock In {\em Database Theory - {ICDT} 2009, 12th International Conference,
  St. Petersburg, Russia, March 23-25, 2009, Proceedings}, pages 31--41, 2009.

\bibitem{DBLP:journals/iandc/ChomickiM05}
Jan Chomicki and Jerzy Marcinkowski.
\newblock Minimal-change integrity maintenance using tuple deletions.
\newblock {\em Inf. Comput.}, 197(1-2):90--121, 2005.

\bibitem{DBLP:conf/pods/FaginKK15}
Ronald Fagin, Benny Kimelfeld, and Phokion~G. Kolaitis.
\newblock Dichotomies in the complexity of preferred repairs.
\newblock In {\em Proceedings of the 34th {ACM} Symposium on Principles of
  Database Systems, {PODS} 2015, Melbourne, Victoria, Australia, May 31 - June
  4, 2015}, pages 3--15, 2015.

\bibitem{DBLP:books/daglib/0005327}
David~R. Brillinger.
\newblock {\em Time series - data analysis and theory}, volume~36 of {\em
  Classics in applied mathematics}.
\newblock {SIAM}, 2001.

\bibitem{DBLP:journals/envsoft/HillM10}
David~J. Hill and Barbara~S. Minsker.
\newblock Anomaly detection in streaming environmental sensor data: {A}
  data-driven modeling approach.
\newblock {\em Environmental Modelling and Software}, 25(9):1014--1022, 2010.

\bibitem{DBLP:conf/kdd/YamanishiT02}
Kenji Yamanishi and Jun'ichi Takeuchi.
\newblock A unifying framework for detecting outliers and change points from
  non-stationary time series data.
\newblock In {\em Proceedings of the Eighth {ACM} {SIGKDD} International
  Conference on Knowledge Discovery and Data Mining, July 23-26, 2002,
  Edmonton, Alberta, Canada}, pages 676--681, 2002.

\bibitem{dilling2017cleaning}
S~Dilling and BJ~MacVicar.
\newblock Cleaning high-frequency velocity profile data with autoregressive
  moving average (arma) models.
\newblock {\em Flow Measurement and Instrumentation}, 54:68--81, 2017.

\bibitem{DBLP:conf/icassp/AlengrinF78}
G{\'{e}}rard Alengrin and G{\'{e}}rard Favier.
\newblock New stochastic realization algorithms for identification of {ARMA}
  models.
\newblock In {\em {IEEE} International Conference on Acoustics, Speech, and
  Signal Processing, {ICASSP} '78, Tulsa, Oklahoma, USA, April 10-12, 1978},
  pages 208--213, 1978.

\bibitem{kalman1960new}
Rudolph~Emil Kalman.
\newblock A new approach to linear filtering and prediction problems.
\newblock {\em Journal of basic Engineering}, 82(1):35--45, 1960.

\bibitem{marczak2018data}
Martyna Marczak, Tommaso Proietti, and Stefano Grassi.
\newblock A data-cleaning augmented kalman filter for robust estimation of
  state space models.
\newblock {\em Econometrics and Statistics}, 5:107--123, 2018.

\bibitem{morrison1977kalman}
G~Wayne Morrison and David~H Pike.
\newblock Kalman filtering applied to statistical forecasting.
\newblock {\em Management Science}, 23(7):768--774, 1977.

\bibitem{brown1992introduction}
Robert~Grover Brown, Patrick~YC Hwang, et~al.
\newblock {\em Introduction to random signals and applied Kalman filtering},
  volume~3.
\newblock Wiley New York, 1992.

\bibitem{DBLP:journals/tsp/EinickeW99}
Garry~A. Einicke and Langford~B. White.
\newblock Robust extended kalman filtering.
\newblock {\em {IEEE} Trans. Signal Processing}, 47(9):2596--2599, 1999.

\bibitem{DBLP:journals/taslp/GohTT99}
Zenton Goh, Kah{-}Chye Tan, and B.~T.~G. Tan.
\newblock Kalman-filtering speech enhancement method based on a voiced-unvoiced
  speech model.
\newblock {\em {IEEE} Trans. Speech and Audio Processing}, 7(5):510--524, 1999.

\bibitem{DBLP:conf/icdcs/Zhuang0WL07}
Yongzhen Zhuang, Lei Chen, Xiaoyang~Sean Wang, and Jie Lian.
\newblock A weighted moving average-based approach for cleaning sensor data.
\newblock In {\em 27th {IEEE} International Conference on Distributed Computing
  Systems {(ICDCS} 2007), June 25-29, 2007, Toronto, Ontario, Canada}, page~38,
  2007.

\bibitem{gardner2006exponential}
Everette~S Gardner~Jr.
\newblock Exponential smoothing: The state of the art---part ii.
\newblock {\em International journal of forecasting}, 22(4):637--666, 2006.

\bibitem{DBLP:conf/icdm/KeoghCHP01}
Eamonn~J. Keogh, Selina Chu, David~M. Hart, and Michael~J. Pazzani.
\newblock An online algorithm for segmenting time series.
\newblock In {\em Proceedings of the 2001 {IEEE} International Conference on
  Data Mining, 29 November - 2 December 2001, San Jose, California, {USA}},
  pages 289--296, 2001.

\bibitem{xu2015data}
Shu Xu, Bo~Lu, Michael Baldea, Thomas~F Edgar, Willy Wojsznis, Terrence
  Blevins, and Mark Nixon.
\newblock Data cleaning in the process industries.
\newblock {\em Reviews in Chemical Engineering}, 31(5):453--490, 2015.

\bibitem{jones1966exponential}
Richard~H Jones.
\newblock Exponential smoothing for multivariate time series.
\newblock {\em Journal of the Royal Statistical Society: Series B
  (Methodological)}, 28(1):241--251, 1966.

\bibitem{van2005accurate}
JWC Van~Lint, SP~Hoogendoorn, and Henk~J van Zuylen.
\newblock Accurate freeway travel time prediction with state-space neural
  networks under missing data.
\newblock {\em Transportation Research Part C: Emerging Technologies},
  13(5-6):347--369, 2005.

\bibitem{DBLP:journals/tip/ChenXF12}
Minjie Chen, Mantao Xu, and Pasi Fr{\"{a}}nti.
\newblock A fast {\textdollar}o(n){\textdollar} multiresolution polygonal
  approximation algorithm for {GPS} trajectory simplification.
\newblock {\em {IEEE} Trans. Image Processing}, 21(5):2770--2785, 2012.

\bibitem{DBLP:journals/pvldb/LongWJ14}
Cheng Long, Raymond~Chi{-}Wing Wong, and H.~V. Jagadish.
\newblock Trajectory simplification: On minimizing the direction-based error.
\newblock {\em {PVLDB}}, 8(1):49--60, 2014.

\bibitem{DBLP:conf/sigmod/DongH82}
Jirun Dong and Richard Hull.
\newblock Applying approximate order dependency to reduce indexing space.
\newblock In {\em Proceedings of the 1982 {ACM} {SIGMOD} International
  Conference on Management of Data, Orlando, Florida, June 2-4, 1982.}, pages
  119--127, 1982.

\bibitem{DBLP:journals/tcs/GinsburgH83}
Seymour Ginsburg and Richard Hull.
\newblock Order dependency in the relational model.
\newblock {\em Theor. Comput. Sci.}, 26:149--195, 1983.

\bibitem{DBLP:conf/pods/GinsburgH83}
Seymour Ginsburg and Richard Hull.
\newblock Sort sets in the relational model.
\newblock In {\em Proceedings of the Second {ACM} {SIGACT-SIGMOD} Symposium on
  Principles of Database Systems, March 21-23, 1983, Colony Square Hotel,
  Atlanta, Georgia, {USA}}, pages 332--339, 1983.

\bibitem{DBLP:journals/jacm/GinsburgH86}
Seymour Ginsburg and Richard Hull.
\newblock Sort sets in the relational model.
\newblock {\em J. ACM}, 33(3):465--488, 1986.

\bibitem{DBLP:conf/icde/LopatenkoB07}
Andrei Lopatenko and Loreto Bravo.
\newblock Efficient approximation algorithms for repairing inconsistent
  databases.
\newblock In {\em Proceedings of the 23rd International Conference on Data
  Engineering, {ICDE} 2007, The Marmara Hotel, Istanbul, Turkey, April 15-20,
  2007}, pages 216--225, 2007.

\bibitem{DBLP:journals/pvldb/GolabKKSS09}
Lukasz Golab, Howard~J. Karloff, Flip Korn, Avishek Saha, and Divesh
  Srivastava.
\newblock Sequential dependencies.
\newblock {\em PVLDB}, 2(1):574--585, 2009.

\bibitem{DBLP:conf/sigmod/SongZWY15}
Shaoxu Song, Aoqian Zhang, Jianmin Wang, and Philip~S. Yu.
\newblock {SCREEN:} stream data cleaning under speed constraints.
\newblock In {\em Proceedings of the 2015 {ACM} {SIGMOD} International
  Conference on Management of Data, Melbourne, Victoria, Australia, May 31 -
  June 4, 2015}, pages 827--841, 2015.

\bibitem{DBLP:conf/dasfaa/YinYWHL18}
Wei Yin, Tianbai Yue, Hongzhi Wang, Yanhao Huang, and Yaping Li.
\newblock Time series cleaning under variance constraints.
\newblock In {\em Database Systems for Advanced Applications - {DASFAA} 2018
  International Workshops: BDMS, BDQM, GDMA, and SeCoP, Gold Coast, QLD,
  Australia, May 21-24, 2018, Proceedings}, pages 108--113, 2018.

\bibitem{DBLP:journals/tkde/SongSZCW20}
Shaoxu Song, Yu~Sun, Aoqian Zhang, Lei Chen, and Jianmin Wang.
\newblock Enriching data imputation under similarity rule constraints.
\newblock {\em {IEEE} Trans. Knowl. Data Eng.}, 32(2):275--287, 2020.

\bibitem{DBLP:conf/icde/ZhangSSW19}
Aoqian Zhang, Shaoxu Song, Yu~Sun, and Jianmin Wang.
\newblock Learning individual models for imputation.
\newblock In {\em 35th {IEEE} International Conference on Data Engineering,
  {ICDE} 2019, Macao, China, April 8-11, 2019}, pages 160--171, 2019.

\bibitem{DBLP:journals/pvldb/CaiX0JOZ18}
Qingchao Cai, Zhongle Xie, Gang Chen, H.~V. Jagadish, Beng~Chin Ooi, and Meihui
  Zhang.
\newblock Effective temporal dependence discovery in time series data.
\newblock {\em {PVLDB}}, 11(8):893--905, 2018.

\bibitem{DBLP:conf/kdd/ChengB014}
Dehua Cheng, Mohammad~Taha Bahadori, and Yan Liu.
\newblock {FBLG:} a simple and effective approach for temporal dependence
  discovery from time series data.
\newblock In {\em The 20th {ACM} {SIGKDD} International Conference on Knowledge
  Discovery and Data Mining, {KDD} '14, New York, NY, {USA} - August 24 - 27,
  2014}, pages 382--391, 2014.

\bibitem{DBLP:journals/tsp/BreslerM86}
Yoram Bresler and Albert Macovski.
\newblock Exact maximum likelihood parameter estimation of superimposed
  exponential signals in noise.
\newblock {\em {IEEE} Trans. Acoustics, Speech, and Signal Processing},
  34(5):1081--1089, 1986.

\bibitem{DBLP:conf/icdt/GogaczT17}
Tomasz Gogacz and Szymon Toru{\'{n}}czyk.
\newblock Entropy bounds for conjunctive queries with functional dependencies.
\newblock In {\em 20th International Conference on Database Theory, {ICDT}
  2017, March 21-24, 2017, Venice, Italy}, pages 15:1--15:17, 2017.

\bibitem{DBLP:journals/tosn/WangKA14}
Dong Wang, Lance~M. Kaplan, and Tarek~F. Abdelzaher.
\newblock Maximum likelihood analysis of conflicting observations in social
  sensing.
\newblock {\em {TOSN}}, 10(2):30:1--30:27, 2014.

\bibitem{DBLP:conf/sigmod/YakoutBE13}
Mohamed Yakout, Laure Berti{-}{\'{E}}quille, and Ahmed~K. Elmagarmid.
\newblock Don't be scared: use scalable automatic repairing with maximal
  likelihood and bounded changes.
\newblock In {\em Proceedings of the {ACM} {SIGMOD} International Conference on
  Management of Data, {SIGMOD} 2013, New York, NY, USA, June 22-27, 2013},
  pages 553--564, 2013.

\bibitem{DBLP:conf/sigmod/ZhangSW16}
Aoqian Zhang, Shaoxu Song, and Jianmin Wang.
\newblock Sequential data cleaning: {A} statistical approach.
\newblock In {\em Proceedings of the 2016 International Conference on
  Management of Data, {SIGMOD} Conference 2016, San Francisco, CA, USA, June 26
  - July 01, 2016}, pages 909--924, 2016.

\bibitem{DBLP:conf/sigmod/WangDM15}
Xiaolan Wang, Xin~Luna Dong, and Alexandra Meliou.
\newblock Data x-ray: {A} diagnostic tool for data errors.
\newblock In {\em Proceedings of the 2015 {ACM} {SIGMOD} International
  Conference on Management of Data, Melbourne, Victoria, Australia, May 31 -
  June 4, 2015}, pages 1231--1245, 2015.

\bibitem{DBLP:conf/icml/GetoorFKT01}
Lise Getoor, Nir Friedman, Daphne Koller, and Benjamin Taskar.
\newblock Learning probabilistic models of relational structure.
\newblock In {\em Proceedings of the Eighteenth International Conference on
  Machine Learning {(ICML} 2001), Williams College, Williamstown, MA, USA, June
  28 - July 1, 2001}, pages 170--177, 2001.

\bibitem{dukhovny1900markov}
AM~Dukhovny.
\newblock Markov chains with quasitoeplitz transition matrix: Applications.
\newblock {\em International Journal of Stochastic Analysis}, 3(2):141--152,
  1900.

\bibitem{cai1994markov}
Jun Cai.
\newblock A markov model of switching-regime arch.
\newblock {\em Journal of Business \& Economic Statistics}, 12(3):309--316,
  1994.

\bibitem{DBLP:journals/tcom/ZhangK99}
Qinqing Zhang and Saleem~A. Kassam.
\newblock Finite-state markov model for rayleigh fading channels.
\newblock {\em {IEEE} Trans. Communications}, 47(11):1688--1692, 1999.

\bibitem{DBLP:journals/eswa/HassanNK07}
Md.~Rafiul Hassan, Baikunth Nath, and Michael Kirley.
\newblock A fusion model of hmm, {ANN} and {GA} for stock market forecasting.
\newblock {\em Expert Syst. Appl.}, 33(1):171--180, 2007.

\bibitem{DBLP:journals/eswa/DongYKHEK09}
Ming Dong, Dong Yang, Yan Kuang, David He, Serap Erdal, and Donna Kenski.
\newblock Pm\({}_{\mbox{2.5}}\) concentration prediction using hidden
  semi-markov model-based times series data mining.
\newblock {\em Expert Syst. Appl.}, 36(5):9046--9055, 2009.

\bibitem{gupta2012stock}
Aditya Gupta and Bhuwan Dhingra.
\newblock Stock market prediction using hidden markov models.
\newblock In {\em 2012 Students Conference on Engineering and Systems}, pages
  1--4. IEEE, 2012.

\bibitem{DBLP:conf/sigmod/BabaJLPKX16}
Asif~Iqbal Baba, Manfred Jaeger, Hua Lu, Torben~Bach Pedersen, Wei{-}Shinn Ku,
  and Xike Xie.
\newblock Learning-based cleansing for indoor {RFID} data.
\newblock In {\em Proceedings of the 2016 International Conference on
  Management of Data, {SIGMOD} Conference 2016, San Francisco, CA, USA, June 26
  - July 01, 2016}, pages 925--936, 2016.

\bibitem{leema2011effective}
A~Anny Leema and M~Hemalatha.
\newblock An effective and adaptive data cleaning technique for colossal rfid
  data sets in healthcare.
\newblock {\em WSEAS transactions on Information Science and Applications},
  8(6):243--252, 2011.

\bibitem{DBLP:journals/ijguc/XuDLSW18}
He~Xu, Jie Ding, Peng Li, Daniele Sgandurra, and Ruchuan Wang.
\newblock An improved {SMURF} scheme for cleaning {RFID} data.
\newblock {\em {IJGUC}}, 9(2):170--178, 2018.

\bibitem{DBLP:conf/vldb/JefferyGF06}
Shawn~R. Jeffery, Minos~N. Garofalakis, and Michael~J. Franklin.
\newblock Adaptive cleaning for {RFID} data streams.
\newblock In {\em Proceedings of the 32nd International Conference on Very
  Large Data Bases, Seoul, Korea, September 12-15, 2006}, pages 163--174, 2006.

\bibitem{DBLP:conf/icde/ZhengMC19}
Mostafa Milani, Zheng Zheng, and Fei Chiang.
\newblock Currentclean: Spatio-temporal cleaning of stale data.
\newblock In {\em 35th {IEEE} International Conference on Data Engineering,
  {ICDE} 2019, Macao, China, April 8-11, 2019}, pages 172--183, 2019.

\bibitem{shumway1982approach}
Robert~H Shumway and David~S Stoffer.
\newblock An approach to time series smoothing and forecasting using the em
  algorithm.
\newblock {\em Journal of time series analysis}, 3(4):253--264, 1982.

\bibitem{DBLP:conf/sigmod/BergmanMNT15}
Moria Bergman, Tova Milo, Slava Novgorodov, and Wang~Chiew Tan.
\newblock Query-oriented data cleaning with oracles.
\newblock In {\em Proceedings of the 2015 {ACM} {SIGMOD} International
  Conference on Management of Data, Melbourne, Victoria, Australia, May 31 -
  June 4, 2015}, pages 1199--1214, 2015.

\bibitem{DBLP:journals/jmlr/NevilleJ07}
Jennifer Neville and David~D. Jensen.
\newblock Relational dependency networks.
\newblock {\em J. Mach. Learn. Res.}, 8:653--692, 2007.

\bibitem{DBLP:conf/sigmod/MayfieldNP10}
Chris Mayfield, Jennifer Neville, and Sunil Prabhakar.
\newblock {ERACER:} a database approach for statistical inference and data
  cleaning.
\newblock In {\em Proceedings of the {ACM} {SIGMOD} International Conference on
  Management of Data, {SIGMOD} 2010, Indianapolis, Indiana, USA, June 6-10,
  2010}, pages 75--86, 2010.

\bibitem{DBLP:conf/cyberc/DiaoLMYH15}
Yinglong Diao, Ke{-}yan Liu, Xiaoli Meng, Xueshun Ye, and Kaiyuan He.
\newblock A big data online cleaning algorithm based on dynamic outlier
  detection.
\newblock In {\em 2015 International Conference on Cyber-Enabled Distributed
  Computing and Knowledge Discovery, CyberC 2015, Xi'an, China, September
  17-19, 2015}, pages 230--234, 2015.

\bibitem{DBLP:journals/bdr/CorizzoCJ19}
Roberto Corizzo, Michelangelo Ceci, and Nathalie Japkowicz.
\newblock Anomaly detection and repair for accurate predictions in
  geo-distributed big data.
\newblock {\em Big Data Res.}, 16:18--35, 2019.

\bibitem{DBLP:journals/kais/KeoghLLH07}
Eamonn~J. Keogh, Jessica Lin, Sang{-}Hee Lee, and Helga~Van Herle.
\newblock Finding the most unusual time series subsequence: algorithms and
  applications.
\newblock {\em Knowl. Inf. Syst.}, 11(1):1--27, 2007.

\bibitem{DBLP:journals/gpem/GonzalezD03}
Fabio~A. Gonz{\'{a}}lez and Dipankar Dasgupta.
\newblock Anomaly detection using real-valued negative selection.
\newblock {\em Genetic Programming and Evolvable Machines}, 4(4):383--403,
  2003.

\bibitem{dasgupta2002anomaly}
Dipankar Dasgupta and Nivedita~Sumi Majumdar.
\newblock Anomaly detection in multidimensional data using negative selection
  algorithm.
\newblock In {\em Proceedings of the 2002 Congress on Evolutionary Computation.
  CEC'02 (Cat. No. 02TH8600)}, volume~2, pages 1039--1044. IEEE, 2002.

\bibitem{DBLP:conf/icde/AggarwalY15}
Charu~C. Aggarwal and Philip~S. Yu.
\newblock On historical diagnosis of sensor streams.
\newblock In {\em 31st {IEEE} International Conference on Data Engineering,
  {ICDE} 2015, Seoul, South Korea, April 13-17, 2015}, pages 185--194, 2015.

\bibitem{DBLP:conf/icde/KontakiGPTM11}
Maria Kontaki, Anastasios Gounaris, Apostolos~N. Papadopoulos, Kostas Tsichlas,
  and Yannis Manolopoulos.
\newblock Continuous monitoring of distance-based outliers over data streams.
\newblock In {\em Proceedings of the 27th International Conference on Data
  Engineering, {ICDE} 2011, April 11-16, 2011, Hannover, Germany}, pages
  135--146, 2011.

\bibitem{DBLP:conf/icann/LiCJSGN19}
Dan Li, Dacheng Chen, Baihong Jin, Lei Shi, Jonathan Goh, and See{-}Kiong Ng.
\newblock {MAD-GAN:} multivariate anomaly detection for time series data with
  generative adversarial networks.
\newblock In {\em Artificial Neural Networks and Machine Learning - {ICANN}
  2019: Text and Time Series - 28th International Conference on Artificial
  Neural Networks, Munich, Germany, September 17-19, 2019, Proceedings, Part
  {IV}}, pages 703--716, 2019.

\bibitem{DBLP:conf/itsc/SunPLS18}
Yuqiang Sun, Lei Peng, Huiyun Li, and Min Sun.
\newblock Exploration on spatiotemporal data repairing of parking lots based on
  recurrent gans.
\newblock In {\em 21st International Conference on Intelligent Transportation
  Systems, {ITSC} 2018, Maui, HI, USA, November 4-7, 2018}, pages 467--472,
  2018.

\bibitem{DBLP:conf/cikm/FangSCG19}
Chenguang Fang, Shaoxu Song, Zhiwei Chen, and Acan Gui.
\newblock Fine-grained fuel consumption prediction.
\newblock In {\em Proceedings of the 28th {ACM} International Conference on
  Information and Knowledge Management, {CIKM} 2019, Beijing, China, November
  3-7, 2019}, pages 2783--2791, 2019.

\bibitem{DBLP:journals/corr/FilonovLV16}
Pavel Filonov, Andrey Lavrentyev, and Artem Vorontsov.
\newblock Multivariate industrial time series with cyber-attack simulation:
  Fault detection using an lstm-based predictive data model.
\newblock {\em CoRR}, abs/1612.06676, 2016.

\bibitem{DBLP:journals/corr/MalhotraTRAVAS16}
Pankaj Malhotra, Vishnu TV, Anusha Ramakrishnan, Gaurangi Anand, Lovekesh Vig,
  Puneet Agarwal, and Gautam Shroff.
\newblock Multi-sensor prognostics using an unsupervised health index based on
  {LSTM} encoder-decoder.
\newblock {\em CoRR}, abs/1608.06154, 2016.

\bibitem{DBLP:conf/esann/MalhotraVSA15}
Pankaj Malhotra, Lovekesh Vig, Gautam Shroff, and Puneet Agarwal.
\newblock Long short term memory networks for anomaly detection in time series,
  2015.

\bibitem{wangxi}
Xi~Wang.
\newblock Research on irregular time series data cleaning based on speed.
\newblock Master's thesis, Tsinghua University, 2020.

\bibitem{brown2004smoothing}
Robert~Goodell Brown.
\newblock {\em Smoothing, forecasting and prediction of discrete time series}.
\newblock Courier Corporation, 2004.

\bibitem{holt2004forecasting}
Charles~C Holt.
\newblock Forecasting seasonals and trends by exponentially weighted moving
  averages.
\newblock {\em International journal of forecasting}, 20(1):5--10, 2004.

\bibitem{DBLP:journals/tits/ZhangYZHL13}
Zhaosheng Zhang, Diange Yang, Tao Zhang, Qiaochu He, and Xiaomin Lian.
\newblock A study on the method for cleaning and repairing the probe vehicle
  data.
\newblock {\em {IEEE} Trans. Intelligent Transportation Systems},
  14(1):419--427, 2013.

\bibitem{qu2016data}
Z~Qu, Y~Wang, Chong Wang, Nan Qu, and Jia Yan.
\newblock A data cleaning model for electric power big data based on spark
  framework.
\newblock {\em International Journal of Database Theory and Application},
  9(3):137--150, 2016.

\bibitem{DBLP:conf/icde/VolkovsCSM14}
Maksims Volkovs, Fei Chiang, Jaroslaw Szlichta, and Ren{\'{e}}e~J. Miller.
\newblock Continuous data cleaning.
\newblock In {\em {IEEE} 30th International Conference on Data Engineering,
  Chicago, {ICDE} 2014, IL, USA, March 31 - April 4, 2014}, pages 244--255,
  2014.

\bibitem{park2005outlier}
Gyuhae Park, Amanda~C Rutherford, Hoon Sohn, and Charles~R Farrar.
\newblock An outlier analysis framework for impedance-based structural health
  monitoring.
\newblock {\em Journal of Sound and Vibration}, 286(1-2):229--250, 2005.

\bibitem{box1970distribution}
George~EP Box and David~A Pierce.
\newblock Distribution of residual autocorrelations in
  autoregressive-integrated moving average time series models.
\newblock {\em Journal of the American statistical Association},
  65(332):1509--1526, 1970.

\bibitem{akouemo2017data}
Hermine~N Akouemo and Richard~J Povinelli.
\newblock Data improving in time series using arx and ann models.
\newblock {\em IEEE Transactions on Power Systems}, 32(5):3352--3359, 2017.

\bibitem{chen2009arima}
Peiyuan Chen, Troels Pedersen, Birgitte Bak-Jensen, and Zhe Chen.
\newblock Arima-based time series model of stochastic wind power generation.
\newblock {\em IEEE transactions on power systems}, 25(2):667--676, 2009.

\bibitem{ma2004predict}
Jie Ma and Jian-fu Teng.
\newblock Predict chaotic time-series using unscented kalman filter.
\newblock In {\em Proceedings of 2004 International Conference on Machine
  Learning and Cybernetics (IEEE Cat. No. 04EX826)}, volume~2, pages 687--690.
  IEEE, 2004.

\bibitem{chang1988estimation}
Ih~Chang, George~C Tiao, and Chung Chen.
\newblock Estimation of time series parameters in the presence of outliers.
\newblock {\em Technometrics}, 30(2):193--204, 1988.

\bibitem{DBLP:journals/tsp/SwamiM90}
Ananthram Swami and Jerry~M. Mendel.
\newblock {ARMA} parameter estimation using only output cumulants.
\newblock {\em {IEEE} Trans. Acoustics, Speech, and Signal Processing},
  38(7):1257--1265, 1990.

\bibitem{plett2004extended}
Gregory~L Plett.
\newblock Extended kalman filtering for battery management systems of
  lipb-based hev battery packs: Part 3. state and parameter estimation.
\newblock {\em Journal of Power sources}, 134(2):277--292, 2004.

\bibitem{DBLP:journals/is/Wijsen98}
Jef Wijsen.
\newblock Reasoning about qualitative trends in databases.
\newblock {\em Inf. Syst.}, 23(7):463--487, 1998.

\bibitem{DBLP:journals/tkde/Wijsen01}
Jef Wijsen.
\newblock Trends in databases: Reasoning and mining.
\newblock {\em IEEE Trans. Knowl. Data Eng.}, 13(3):426--438, 2001.

\bibitem{DBLP:conf/sigmod/SongZW16}
Shaoxu Song, Han Zhu, and Jianmin Wang.
\newblock Constraint-variance tolerant data repairing.
\newblock In {\em Proceedings of the 2016 International Conference on
  Management of Data, {SIGMOD} Conference 2016, San Francisco, CA, USA, June 26
  - July 01, 2016}, pages 877--892, 2016.

\bibitem{DBLP:conf/sensys/Casado-VaraPPC18}
Roberto Casado{-}Vara, Fernando de~la Prieta, Javier Prieto, and Juan~M.
  Corchado.
\newblock Blockchain framework for iot data quality via edge computing.
\newblock In {\em Proceedings of the 1st Workshop on Blockchain-enabled
  Networked Sensor Systems, BlockSys@SenSys 2018, Shenzhen, China, November 4,
  2018}, pages 19--24, 2018.

\bibitem{beeri1984structure}
Catriel Beeri, Martin Dowd, Ronald Fagin, and Richard Statman.
\newblock On the structure of armstrong relations for functional dependencies.
\newblock {\em Journal of the ACM (JACM)}, 31(1):30--46, 1984.

\bibitem{fan2008conditional}
Wenfei Fan, Floris Geerts, Xibei Jia, and Anastasios Kementsietsidis.
\newblock Conditional functional dependencies for capturing data
  inconsistencies.
\newblock {\em ACM Transactions on Database Systems (TODS)}, 33(2):6, 2008.

\bibitem{DBLP:journals/tkde/FanGLX11}
Wenfei Fan, Floris Geerts, Jianzhong Li, and Ming Xiong.
\newblock Discovering conditional functional dependencies.
\newblock {\em {IEEE} Trans. Knowl. Data Eng.}, 23(5):683--698, 2011.

\bibitem{DBLP:conf/cikm/SongC09}
Shaoxu Song and Lei Chen.
\newblock Discovering matching dependencies.
\newblock In {\em Proceedings of the 18th {ACM} Conference on Information and
  Knowledge Management, {CIKM} 2009, Hong Kong, China, November 2-6, 2009},
  pages 1421--1424, 2009.

\bibitem{DBLP:journals/dke/Song013}
Shaoxu Song and Lei Chen.
\newblock Efficient discovery of similarity constraints for matching
  dependencies.
\newblock {\em Data Knowl. Eng.}, 87:146--166, 2013.

\bibitem{DBLP:journals/tkdd/WangSCYC17}
Yihan Wang, Shaoxu Song, Lei Chen, Jeffrey~Xu Yu, and Hong Cheng.
\newblock Discovering conditional matching rules.
\newblock {\em {TKDD}}, 11(4):46:1--46:38, 2017.

\bibitem{DBLP:journals/tods/Song011}
Shaoxu Song and Lei Chen.
\newblock Differential dependencies: Reasoning and discovery.
\newblock {\em {ACM} Trans. Database Syst.}, 36(3):16:1--16:41, 2011.

\bibitem{DBLP:conf/icde/SongCC12}
Shaoxu Song, Lei Chen, and Hong Cheng.
\newblock Parameter-free determination of distance thresholds for metric
  distance constraints.
\newblock In {\em {IEEE} 28th International Conference on Data Engineering
  {(ICDE} 2012), Washington, DC, {USA} (Arlington, Virginia), 1-5 April, 2012},
  pages 846--857, 2012.

\bibitem{DBLP:journals/tkde/Song0C14}
Shaoxu Song, Lei Chen, and Hong Cheng.
\newblock Efficient determination of distance thresholds for differential
  dependencies.
\newblock {\em {IEEE} Trans. Knowl. Data Eng.}, 26(9):2179--2192, 2014.

\bibitem{DBLP:conf/icde/SongCY11}
Shaoxu Song, Lei Chen, and Philip~S. Yu.
\newblock On data dependencies in dataspaces.
\newblock In {\em Proceedings of the 27th International Conference on Data
  Engineering, {ICDE} 2011, April 11-16, 2011, Hannover, Germany}, pages
  470--481, 2011.

\bibitem{DBLP:journals/vldb/Song0Y13}
Shaoxu Song, Lei Chen, and Philip~S. Yu.
\newblock Comparable dependencies over heterogeneous data.
\newblock {\em {VLDB} J.}, 22(2):253--274, 2013.

\bibitem{DBLP:journals/isci/SongZ014}
Shaoxu Song, Han Zhu, and Lei Chen.
\newblock Probabilistic correlation-based similarity measure on text records.
\newblock {\em Inf. Sci.}, 289:8--24, 2014.

\bibitem{DBLP:journals/pvldb/SongCY014}
Shaoxu Song, Hong Cheng, Jeffrey~Xu Yu, and Lei Chen.
\newblock Repairing vertex labels under neighborhood constraints.
\newblock {\em {PVLDB}}, 7(11):987--998, 2014.

\bibitem{DBLP:journals/vldb/SongLCYC17}
Shaoxu Song, Boge Liu, Hong Cheng, Jeffrey~Xu Yu, and Lei Chen.
\newblock Graph repairing under neighborhood constraints.
\newblock {\em {VLDB} J.}, 26(5):611--635, 2017.

\bibitem{DBLP:journals/pvldb/SongZC015}
Shaoxu Song, Aoqian Zhang, Lei Chen, and Jianmin Wang.
\newblock Enriching data imputation with extensive similarity neighbors.
\newblock {\em {PVLDB}}, 8(11):1286--1297, 2015.

\bibitem{DBLP:conf/sigmod/ZhouT15}
Jingbo Zhou and Anthony K.~H. Tung.
\newblock Smiler: {A} semi-lazy time series prediction system for sensors.
\newblock In {\em Proceedings of the 2015 {ACM} {SIGMOD} International
  Conference on Management of Data, Melbourne, Victoria, Australia, May 31 -
  June 4, 2015}, pages 1871--1886, 2015.

\bibitem{DBLP:journals/tii/TianZDHSCWW19}
Daxin Tian, Yukai Zhu, Xuting Duan, Junjie Hu, Zhengguo Sheng, Min Chen, Jian
  Wang, and Yunpeng Wang.
\newblock An effective fuel-level data cleaning and repairing method for
  vehicle monitor platform.
\newblock {\em {IEEE} Trans. Industrial Informatics}, 15(1):410--422, 2019.

\bibitem{dempster1977maximum}
Arthur~P Dempster, Nan~M Laird, and Donald~B Rubin.
\newblock Maximum likelihood from incomplete data via the em algorithm.
\newblock {\em Journal of the Royal Statistical Society: Series B
  (Methodological)}, 39(1):1--22, 1977.

\bibitem{DBLP:journals/kais/BasuM07}
Sabyasachi Basu and Martin Meckesheimer.
\newblock Automatic outlier detection for time series: an application to sensor
  data.
\newblock {\em Knowl. Inf. Syst.}, 11(2):137--154, 2007.

\bibitem{DBLP:journals/ijon/Zhang03}
Guoqiang~Peter Zhang.
\newblock Time series forecasting using a hybrid {ARIMA} and neural network
  model.
\newblock {\em Neurocomputing}, 50:159--175, 2003.

\bibitem{DBLP:conf/icdm/KeoghLF05}
Eamonn~J. Keogh, Jessica Lin, and Ada~Wai{-}Chee Fu.
\newblock {HOT} {SAX:} efficiently finding the most unusual time series
  subsequence.
\newblock In {\em Proceedings of the 5th {IEEE} International Conference on
  Data Mining {(ICDM} 2005), 27-30 November 2005, Houston, Texas, {USA}}, pages
  226--233, 2005.

\bibitem{DBLP:conf/icdm/WeiKX06}
Li~Wei, Eamonn~J. Keogh, and Xiaopeng Xi.
\newblock Saxually explicit images: Finding unusual shapes.
\newblock In {\em Proceedings of the 6th {IEEE} International Conference on
  Data Mining {(ICDM} 2006), 18-22 December 2006, Hong Kong, China}, pages
  711--720, 2006.

\bibitem{DBLP:conf/kdd/KeoghLR04}
Eamonn~J. Keogh, Stefano Lonardi, and Chotirat~(Ann) Ratanamahatana.
\newblock Towards parameter-free data mining.
\newblock In {\em Proceedings of the Tenth {ACM} {SIGKDD} International
  Conference on Knowledge Discovery and Data Mining, Seattle, Washington, USA,
  August 22-25, 2004}, pages 206--215, 2004.

\bibitem{DBLP:conf/id/GhoshSS99}
Anup~K. Ghosh, Aaron Schwartzbard, and Michael Schatz.
\newblock Learning program behavior profiles for intrusion detection.
\newblock In {\em Proceedings of the Workshop on Intrusion Detection and
  Network Monitoring, Santa Clara, CA, USA, April 9-12, 1999}, pages 51--62,
  1999.

\bibitem{DBLP:conf/acsac/Endler98}
D.~Endler.
\newblock Intrusion detection applying machine learning to solaris audit data.
\newblock In {\em 14th Annual Computer Security Applications Conference
  {(ACSAC} 1998), 7-11 December 1998, Scottsdale, AZ, {USA}}, pages 268--279,
  1998.

\bibitem{DBLP:conf/edbt/Senin0WOGBCF15}
Pavel Senin, Jessica Lin, Xing Wang, Tim Oates, Sunil Gandhi, Arnold~P.
  Boedihardjo, Crystal Chen, and Susan Frankenstein.
\newblock Time series anomaly discovery with grammar-based compression.
\newblock In {\em Proceedings of the 18th International Conference on Extending
  Database Technology, {EDBT} 2015, Brussels, Belgium, March 23-27, 2015},
  pages 481--492, 2015.

\bibitem{DBLP:conf/kdd/EsterKSX96}
Martin Ester, Hans{-}Peter Kriegel, J{\"{o}}rg Sander, and Xiaowei Xu.
\newblock A density-based algorithm for discovering clusters in large spatial
  databases with noise.
\newblock In {\em Proceedings of the Second International Conference on
  Knowledge Discovery and Data Mining (KDD-96), Portland, Oregon, {USA}}, pages
  226--231, 1996.

\bibitem{DBLP:conf/kdd/SongLZ15}
Shaoxu Song, Chunping Li, and Xiaoquan Zhang.
\newblock Turn waste into wealth: On simultaneous clustering and cleaning over
  dirty data.
\newblock In {\em Proceedings of the 21th {ACM} {SIGKDD} International
  Conference on Knowledge Discovery and Data Mining, Sydney, NSW, Australia,
  August 10-13, 2015}, pages 1115--1124, 2015.

\bibitem{DBLP:journals/neco/GersSC00}
Felix~A. Gers, J{\"{u}}rgen Schmidhuber, and Fred~A. Cummins.
\newblock Learning to forget: Continual prediction with {LSTM}.
\newblock {\em Neural Computation}, 12(10):2451--2471, 2000.

\bibitem{DBLP:journals/corr/MalhotraRAVAS16}
Pankaj Malhotra, Anusha Ramakrishnan, Gaurangi Anand, Lovekesh Vig, Puneet
  Agarwal, and Gautam Shroff.
\newblock Lstm-based encoder-decoder for multi-sensor anomaly detection.
\newblock {\em CoRR}, abs/1607.00148, 2016.

\bibitem{DBLP:journals/kais/XingPY12}
Zhengzheng Xing, Jian Pei, and Philip~S. Yu.
\newblock Early classification on time series.
\newblock {\em Knowl. Inf. Syst.}, 31(1):105--127, 2012.

\bibitem{DBLP:conf/sigmod/BreunigKNS00}
Markus~M. Breunig, Hans{-}Peter Kriegel, Raymond~T. Ng, and J{\"{o}}rg Sander.
\newblock {LOF:} identifying density-based local outliers.
\newblock In {\em Proceedings of the 2000 {ACM} {SIGMOD} International
  Conference on Management of Data, May 16-18, 2000, Dallas, Texas, {USA.}},
  pages 93--104, 2000.

\bibitem{DBLP:journals/pvldb/ZhangS0Y17}
Aoqian Zhang, Shaoxu Song, Jianmin Wang, and Philip~S. Yu.
\newblock Time series data cleaning: From anomaly detection to anomaly
  repairing.
\newblock {\em {PVLDB}}, 10(10):1046--1057, 2017.

\bibitem{DBLP:conf/sigmod/YuC19}
Zhuoran Yu and Xu~Chu.
\newblock Piclean: {A} probabilistic and interactive data cleaning system.
\newblock In {\em Proceedings of the 2019 International Conference on
  Management of Data, {SIGMOD} Conference 2019, Amsterdam, The Netherlands,
  June 30 - July 5, 2019.}, pages 2021--2024, 2019.

\bibitem{DBLP:journals/pvldb/RekatsinasCIR17}
Theodoros Rekatsinas, Xu~Chu, Ihab~F. Ilyas, and Christopher R{\'{e}}.
\newblock Holoclean: Holistic data repairs with probabilistic inference.
\newblock {\em {PVLDB}}, 10(11):1190--1201, 2017.

\bibitem{DBLP:journals/pvldb/KrishnanWWFG16}
Sanjay Krishnan, Jiannan Wang, Eugene Wu, Michael~J. Franklin, and Ken
  Goldberg.
\newblock Activeclean: Interactive data cleaning for statistical modeling.
\newblock {\em {PVLDB}}, 9(12):948--959, 2016.

\bibitem{DBLP:journals/pvldb/DingWSLLG19}
Xiaoou Ding, Hongzhi Wang, Jiaxuan Su, Zijue Li, Jianzhong Li, and Hong Gao.
\newblock Cleanits: {A} data cleaning system for industrial time series.
\newblock {\em {PVLDB}}, 12(12):1786--1789, 2019.

\bibitem{DBLP:conf/sigmod/TaeROKW19}
Ki~Hyun Tae, Yuji Roh, Young~Hun Oh, Hyunsu Kim, and Steven~Euijong Whang.
\newblock Data cleaning for accurate, fair, and robust models: {A} big data -
  {AI} integration approach.
\newblock In {\em Proceedings of the 3rd International Workshop on Data
  Management for End-to-End Machine Learning, DEEM@SIGMOD 2019, Amsterdam, The
  Netherlands, June 30, 2019}, pages 5:1--5:4, 2019.

\bibitem{DBLP:journals/pvldb/RongB17}
Kexin Rong and Peter Bailis.
\newblock {ASAP:} prioritizing attention via time series smoothing.
\newblock {\em {PVLDB}}, 10(11):1358--1369, 2017.

\bibitem{DBLP:conf/icc/0004ZFWYY19}
Jinlin Wang, Hongli Zhang, Binxing Fang, Xing Wang, Gongzhu Yin, and Ximiao Yu.
\newblock Edcleaner: Data cleaning for entity information in social network.
\newblock In {\em 2019 {IEEE} International Conference on Communications, {ICC}
  2019, Shanghai, China, May 20-24, 2019}, pages 1--7, 2019.

\bibitem{DBLP:conf/bigdataconf/HuangMC18}
Yu~Huang, Mostafa Milani, and Fei Chiang.
\newblock {PACAS:} privacy-aware, data cleaning-as-a-service.
\newblock In {\em {IEEE} International Conference on Big Data, Big Data 2018,
  Seattle, WA, USA, December 10-13, 2018}, pages 1023--1030, 2018.

\bibitem{DBLP:conf/bigdataconf/HuangCLS019}
Ruihong Huang, Zhiwei Chen, Zhicheng Liu, Shaoxu Song, and Jianmin Wang.
\newblock Tsoutlier: Explaining outliers with uniform profiles over iot data.
\newblock In {\em 2019 {IEEE} International Conference on Big Data (Big Data),
  Los Angeles, CA, USA, December 9-12, 2019}, pages 2024--2027, 2019.

\bibitem{DBLP:conf/kdd/LaptevAF15}
Nikolay Laptev, Saeed Amizadeh, and Ian Flint.
\newblock Generic and scalable framework for automated time-series anomaly
  detection.
\newblock In {\em Proceedings of the 21th {ACM} {SIGKDD} International
  Conference on Knowledge Discovery and Data Mining, Sydney, NSW, Australia,
  August 10-13, 2015}, pages 1939--1947, 2015.

\bibitem{DBLP:journals/pvldb/DasuL12}
Tamraparni Dasu and Ji~Meng Loh.
\newblock Statistical distortion: Consequences of data cleaning.
\newblock {\em {PVLDB}}, 5(11):1674--1683, 2012.

\bibitem{DBLP:journals/pvldb/0001SZL13}
Jianmin Wang, Shaoxu Song, Xiaochen Zhu, and Xuemin Lin.
\newblock Efficient recovery of missing events.
\newblock {\em {PVLDB}}, 6(10):841--852, 2013.

\bibitem{DBLP:journals/tkde/0001SZLS16}
Jianmin Wang, Shaoxu Song, Xiaochen Zhu, Xuemin Lin, and Jiaguang Sun.
\newblock Efficient recovery of missing events.
\newblock {\em {IEEE} Trans. Knowl. Data Eng.}, 28(11):2943--2957, 2016.

\bibitem{DBLP:conf/sigmod/LianCS10}
Xiang Lian, Lei Chen, and Shaoxu Song.
\newblock Consistent query answers in inconsistent probabilistic databases.
\newblock In {\em Proceedings of the {ACM} {SIGMOD} International Conference on
  Management of Data, {SIGMOD} 2010, Indianapolis, Indiana, USA, June 6-10,
  2010}, pages 303--314, 2010.

\bibitem{DBLP:conf/sigmod/ZhuSL0Z14}
Xiaochen Zhu, Shaoxu Song, Xiang Lian, Jianmin Wang, and Lei Zou.
\newblock Matching heterogeneous event data.
\newblock In {\em International Conference on Management of Data, {SIGMOD}
  2014, Snowbird, UT, USA, June 22-27, 2014}, pages 1211--1222. {ACM}, 2014.

\bibitem{DBLP:journals/tkde/GaoSZWLZ18}
Yu~Gao, Shaoxu Song, Xiaochen Zhu, Jianmin Wang, Xiang Lian, and Lei Zou.
\newblock Matching heterogeneous event data.
\newblock {\em {IEEE} Trans. Knowl. Data Eng.}, 30(11):2157--2170, 2018.

\bibitem{DBLP:conf/icde/ZhuSWYS14}
Xiaochen Zhu, Shaoxu Song, Jianmin Wang, Philip~S. Yu, and Jiaguang Sun.
\newblock Matching heterogeneous events with patterns.
\newblock In {\em {IEEE} 30th International Conference on Data Engineering,
  Chicago, {ICDE} 2014, IL, USA, March 31 - April 4, 2014}, pages 376--387,
  2014.

\bibitem{DBLP:journals/tkde/SongGWZWY17}
Shaoxu Song, Yu~Gao, Chaokun Wang, Xiaochen Zhu, Jianmin Wang, and Philip~S.
  Yu.
\newblock Matching heterogeneous events with patterns.
\newblock {\em {IEEE} Trans. Knowl. Data Eng.}, 29(8):1695--1708, 2017.

\end{thebibliography}

\end{document}